\documentclass[prd,twocolumn,showpacs,amsfonts,amsmath,amssymb,superscriptaddress,preprintnumbers,nofootinbib,showkeys]{revtex4}
\usepackage{graphicx}
\usepackage{float}
\graphicspath{{./Plots/}}

\begin{document}
\title{Self-interaction in a cosmic dark fluid: \\ The four-kernel rheological extension of the equations of state}
\author{Alexander B. Balakin}
\email{Alexander.Balakin@kpfu.ru}
\author{Alexei S. Ilin}
\email{alexeyilinjukeu@gmail.com}
\affiliation{Department of General Relativity and
Gravitation, Institute of Physics,Kazan Federal University, Kremlevskaya street 18, Kazan, 420008,
Russia}
\date{\today}

\begin{abstract}
We establish a new self-consistent model of coupling between the cosmic dark energy and dark matter in the framework of the rheological approach, which is based on the representation of the equations of state in terms of integral operators of the Volterra-type. We elaborate the so-called four-kernel model, in the framework of which both the dark energy and dark matter pressures are presented by two integrals containing the energy densities of the dark energy and dark matter. For the Volterra operators, the kernels of which are associated with the effects of fading memory, the corresponding isotropic homogeneous cosmological model is shown to be exactly integrable. We consider the classification of the model exact solutions, based on the analysis of roots of the characteristic polynomial associated with the key equation of the presented model. The scalars of the pressure and energy-density of the dark energy and dark matter, the Hubble function and acceleration parameter are presented explicitly as the functions of the dimensionless scale factor. The scale factor as the function of the cosmological time is found in quadratures and is described analytically, qualitatively and numerically. Asymptotic analysis allowed us to classify the models with respect to behavior typical for the Big Rip, Little Rip and Pseudo Rip (de Sitter type). Two intriguing exact cosmological solutions are discussed, which describe the super-exponential expansion and the symmetric bounce. New solutions are presented, which correspond to the quasi-periodic behavior of the state functions of the dark fluid and of the geometric characteristics of the Universe.
\end{abstract}
\maketitle

\section{Introduction}

\subsection{On the problem of internal interactions in the cosmic dark fluid}

The cosmic dark fluid, which consists of dark matter  and dark energy, plays the key role in all modern cosmological scenaria  \cite{DM1}~-~\cite{DMDE4}. The dark matter and dark energy interact by the gravitational field, thus creating the space-time background for various astrophysical and cosmological events. Observational data, obtained recently,  show that the direct (non-gravitational) interaction between dark matter and dark energy cannot be excluded \cite{IDMDE1,IDMDE2,IDMDE3,IDMDE4}. The concept of non-gravitational interaction in the dark sector, precisely between dark matter (DM) and dark energy (DE), was phenomenologically introduced to explain, in particular, the cosmic coincidence problem \cite{CO1,CO2,CO3}. There are few models of interactions in the dark sector. The most known phenomenological model operates with the so-called kernel of interaction, the function $Q(t)$, which appears in the individual balance equations for the DE and DM energy densities with opposite signs, $+Q$ and $-Q$, thus providing the conservation of the total (DE+DM) energy density (see, e.g., \cite{i1,Z,Pavon1,Pavon2}).
In the series of works \cite{Arc1,Arc2,Arc3,Arc4} the DE/DM interaction is modeled on the base of relativistic kinetic theory with an assumption that DE acts on the DM particles by the gradient force of the Archimedean type. In \cite{BD2014,Obzor} the DE/DM interactions are considered in terms of extended electrodynamics of continua. In \cite{BI2018} the kernel of non-gravitational interaction between DE and DM is presented by the integral Volterra-type operator. The main idea of both differential and integral extensions of the interaction terms is based on the concept that the response of the DM on the DE action (and vice versa) occurs with a time delay, not instantly. This approach is supported by various physical models for the classical matter with rheologic properties, and we hope that the behavior of the dark constituents of the cosmic dark fluid is similar in this sense. As for the self-interaction inside the DE and DM, there are models (see, e.g., \cite{Odin} and \cite{Arc1}), in which the equations of state of the  DE are extended by the terms with the first derivative of the DE pressure. The extension of this type was inspired by the results of the relativistic causal thermodynamics, elaborated by Israel and Stewart \cite{IS}.
In fact, the appearance of the differential and/or  integral terms in the equations of state for DM and DE reveals the intention to describe the simplest effects of nonlocality in time.
One can mention two classical theories, which have realized this paradigm: the theory of viscoelasticity and rheology (see, e.g., \cite{Visco,Rheo,Rabotnov,JCL,Maugin99}). In these theories the concept of fading memory is used, and the corresponding mathematical formalism is based on the theory of linear Volterra operators \cite{Volterra}, which considers the value of the pressure at the moment to be  predetermined by whole prehistory of the material evolution.
Generally, the problem of theoretical description of nonlocal interactions is well known in physics, and particularly, in  cosmology and theories of gravity (see, e.g., \cite{NLG1,NLG2,NLG3,NLG4,NLG5,NLG6}. We intend to involve the formalism of the nonlocal theory to the problem of internal interactions in the cosmic dark fluid, using the isotropic homogeneous spacetime platform.

\subsection{Prologue}

Classical theory of viscoelasticity \cite{Visco} operates with two local constitutive laws, first, with the Hooke law, which states that the stress $\sigma$ is proportional to the strain $\epsilon$, second with the Newton law, which claims that the stress is proportional to the time derivative of the strain $\dot{\epsilon}$. Symbolically, these laws can be written as follows:
\begin{equation}
\sigma = E_0 \epsilon \,, \quad \sigma = \eta \dot{\epsilon}\,,
\end{equation}
where the parameter $E_0$ describes the elastic modulus, and $\eta$ relates to the viscosity coefficient. For the schematic illustration of the material properties one uses combination of springs, which symbolize the Hooke's properties, and of the dashpots, when one deals with the behavior of the Newton type.
Serial connection of one spring and one dashpot symbolizes the Maxwell model of viscoelasticity, which can be described by the constitutive equation
\begin{equation}
\dot{\sigma} + \frac{E_0}{\eta}\sigma = E_0 \dot{\epsilon} \,.
\end{equation}
This constitutive equation can be rewritten in the integral form
\begin{equation}
\sigma(t) = \sigma(0)e^{-\frac{E_0}{\eta} t } + E_0 \int_{0}^{t} d \tau \dot{\epsilon}(\tau) e^{- \frac{E_0}{\eta}(t-\tau)} \,.
\label{P3}
\end{equation}
The right-hand side of this formula contains the so-called Volterra integral with the difference multiplicative kernel
\begin{equation}
K(t{-}\tau) = e^{{-} \frac{E_0}{\eta}(t{-}\tau)} = e^{{-} \frac{E_0}{\eta} t} \cdot e^{ \frac{E_0}{\eta}\tau}
\label{p4}
\end{equation}
which describes the fading memory \cite{Visco,Rabotnov}.

When one depicts two springs and two dashpots connected as two parallel Maxwell details, one obtains the Burgers model with the constitutive equation of the second order in time derivative:
$$
\ddot{\sigma} + \dot{\sigma}\left(\frac{E_1}{\eta_1}{+} \frac{E_2}{\eta_2}\right) + \sigma \frac{E_1 E_2}{\eta_1 \eta_2} = f(\epsilon)\,,
$$
\begin{equation}
f(t) \equiv \dot{\epsilon} E_1 E_2 \left(\frac{1}{\eta_1}+ \frac{1}{\eta_2} \right) + \ddot{\epsilon}(E_1+E_2) \,.
\label{p5}
\end{equation}
This differential relationship is equivalent to the integral one
$$
\sigma(t) = \left[\sigma(0) \cosh{\tilde{\Gamma} t} + \frac{\gamma \sigma(0)+ \dot{\sigma}(0)}{\tilde{\Gamma}} \sinh{\tilde{\Gamma} t}\right]e^{- \tilde{\gamma}t} +
$$
\begin{equation}
+ \frac{1}{2\tilde{\Gamma}} \int_0^{t}d \tau f(\tau) \left[e^{- \frac{E_2}{\eta_2}(t-\tau)} - e^{- \frac{E_1}{\eta_1}(t-\tau)} \right] \,,
\label{p6}
\end{equation}
where the parameters $\tilde{\gamma}$ and $\tilde{\Gamma}$ are given by
\begin{equation}
\tilde{\Gamma}  \equiv \frac12 \left(\frac{E_1}{\eta_1} - \frac{E_2}{\eta_2} \right)\,, \quad \tilde{\gamma} \equiv  \frac12 \left(\frac{E_1}{\eta_1} + \frac{E_2}{\eta_2} \right) \,.
\label{p7}
\end{equation}
When $\sigma(0)=0$ and $\dot{\sigma}(0)=0$, we obtain from (\ref{p6}) the integral form of the constitutive equation for the Burgers model.
Clearly, the Burgers model of viscoelasticity deals with the multiplicative kernels of the Volterra type, which is given by the difference of two Maxwell kernels (\ref{p4}). This illustration gives us the analog and motivation for the four-kernel extension of the model of interaction between the dark energy and dark matter.

\subsection{Structure of the work}

In the presented work we deal with the dark fluid consisting of two dark constituents, and we consider two equations of state (EoS). In the EoS for the dark energy the DE pressure is presented by two Volterra integrals containing the DE energy density scalar and DM energy density scalar, respectively. Similarly, the EoS for the dark matter contains two Volterra integrals. Thus, the model requires to introduce four kernels; our ansatz is that all four kernels describe the fading memory and have the multiplicative form.

The paper is organized as follows. In Section II we describe the formalism. i.e., we present the equations of the gravity field, the equations of state for the DE and DM, and the balance equations. In Section III we derive the integro-differential equations describing the evolution of the isotropic homogeneous Universe, and obtain the so-called key equation, which is the linear differential equation of the Euler type in ordinary derivatives for the DE energy density. Depending on the completeness of the set of phenomenologically introduced coupling parameters, the key equation can be of the sixth, fifth, fourth, third and second order in derivative; we describe all the corresponding schemes of derivation and present the sets of auxiliary coefficients in the Appendix I and Appendix II. In section IV we give the classification of the exact solutions to the key equation based on the analysis of solutions to the characteristic equation associated with the Euler equation; using the asymptotic analysis of the obtained solutions we indicate the cases which correspond to the Universe behavior typical for Big Rip, Little Rip and Pseudo Rip. Section V contains explicit examples of  analytic solutions for the model, which describes the pressureless dark matter and non-locally self-interacting dark energy coupled by the local link. In Section VI we consider exact explicit solutions of the model of the non-local cross-action of DE on DM. Section VII contains discussion and conclusions.

\section{The formalism}

\subsection{Two-fluid representation of the isotropic homogeneous cosmological model}

The master equations for the gravity field obtained from the Hilbert-Einstein action functional have the form
\begin{equation}
R^{ik}-\frac12 g^{ik} R - \Lambda g^{ik} = \kappa	\left[T^{ik}_{(\rm DE)} + T^{ik}_{(\rm DM)} \right]\,,
\label{F1}
\end{equation}
where $R^{ik}$ is the Ricci tensor, $R$ is the Ricci scalar, $\Lambda$ is the cosmological constant. The quantities $T^{ik}_{(\rm DE)}$ and $T^{ik}_{(\rm DM)}$ are the stress-energy tensors of the dark energy and dark matter, respectively.
We assume that the spacetime is described by the line element
\begin{equation}
ds^2=dt^2-a^2(t)\left[dx^2+dy^2+dz^2 \right] \,.
\label{F8}
\end{equation}
Our ansatz is that the DE and DM stress-energy tensors have the form
\begin{equation}
T^{ik}_{(\rm DE)} = W U^i U^k -  P \Delta^{ik} \,,
\quad
T^{ik}_{(\rm DM)} = E U^i U^k - \Pi \Delta^{ik} \,.
\label{F2}
\end{equation}
Here $U^i = \delta^i_0$ is the timelike unit velocity four-vector; $\Delta^{ik} \equiv g^{ik}-U^i U^k$ is the projector.  $W$ and $E$ are the energy density scalars of DE and DM, respectively; $P$ and $\Pi$ describe the corresponding pressure scalars. All the state functions are assumed to be the functions of time only.

The Bianchi identity provides the sum of the DE and DM stress-energy tensors to be
divergence free:
\begin{equation}
\nabla_k\left[T^{ik}_{(\rm DE)} + T^{ik}_{(\rm DM)} \right]=0	 \,.
\label{F5}
\end{equation}
In the isotropic homogeneous spacetime with the metric (\ref{F8}) this equality can be rewritten via two balance equations
\begin{equation}\label{EQ2}
\dot{W}+3H(W+P)=Q \,,
\end{equation}
\begin{equation}\label{EQ3}
\dot{E}+3H(E+\Pi)=-Q \,,
\end{equation}
where $H(t) = \frac{\dot{a}(t)}{a(t)}$ is the Hubble function, the dot denotes the derivative with respect to time. The quantity $Q(t)$ is some auxiliary function of time indicated as the kernel of interaction between the DE and DM. We assume that the kernel of DE/DM interaction is of linear form:
\begin{equation}\label{EQ6}
Q(t) = \omega_{0} H(t) [E(t)-W(t)]\,,
\end{equation}
where $\omega_0$ is a dimensionless phenomenological constant. Due to the symmetry of the model we have only one independent equation describing the gravity field, it has the form
\begin{equation}\label{EQ1}
3H^2 - \Lambda = \kappa \left[W(t)+E(t) \right] \,.
\end{equation}
To solve the set of master equations (\ref{EQ1}), (\ref{EQ2}),(\ref{EQ3}) with (\ref{EQ6})
we have to add two equations of state for the DE and DM, respectively:
\begin{equation} \label{EQ4}
P=P(W, E) \,, \quad \Pi=\Pi(W, E) \,.
\end{equation}

\subsection{Reconstruction of the constitutive equations}

We suggest to formulate the equations of state for the dark energy in the following integral form:
$$
P(t) = (\Gamma - 1) W(t) +
$$
\begin{equation}\label{EoS1}
+ \int\limits^t_{t_{0}} d\xi K_{11}(t,\xi) W(\xi) + \int\limits^t_{t_{0}} d\xi K_{12}(t,\xi) E(\xi) \,,
\end{equation}
using two Volterra type operators. Similarly, the equation of state for the dark matter is presented in the form
$$
\Pi(t) = (\gamma - 1) E(t) +
$$
\begin{equation}\label{EoS2}
+ \int\limits^t_{t_{0}} d\xi K_{21}(t,\xi) W(\xi) + \int\limits^t_{t_{0}} d\xi K_{22}(t,\xi) E(\xi) \,.
\end{equation}
These constitutive laws require the following comments.

\noindent
1) When $K_{11}{=}K_{12}{=}K_{21}{=}K_{22}{=}0$ the constitutive laws (\ref{EoS1}) and (\ref{EoS2}) give the standard barotropic equations of state $P(t) = (\Gamma {-}1) W(t)$ and  $\Pi(t) = (\gamma {-} 1) E(t)$, thus, the constants $\Gamma$ and $\gamma$ play the roles of the adiabatic parameters for the DE and DM, respectively.

\noindent
2) When the cross-terms vanish, i.e., $K_{12}{=}K_{21}{=}0$, we deal with two integral type equations of state, which are independent for DE and DM; there are no internal cross-interactions in the dark fluid, but there exist self-interactions in DE and DM individually.

\noindent
3) Generally, $K_{12} \neq K_{21}$, though the symmetric case $K_{12} {=} K_{21}$ is also interesting.

\noindent
4) The constitutive equations (\ref{EoS1}) and (\ref{EoS2}) belong to the class of non-local laws, i.e., the value of the DE and DM pressures at the time moment  $t$ are predetermined by all prehistory of the dark fluid evolution.

\subsection{The Volterra kernels describing the fading memory}

Keeping in mind the classical analogs from the Maxwell and Burgers models of viscoelasticity, we suggest to use the multiplicative Volterra kernels of the following form
\begin{equation}
K_{ij}(t,\xi) = K_{ij}^0 \ H(\xi) \left[\frac{a(\xi)}{a(t)}\right]^{\nu_{ij}} \,,
\label{K1}
\end{equation}
where $i,j = 1,2$; as for the quantities $K_{ij}^0$ and $\nu_{ij}$, they are some dimensionless phenomenological constants. The signs of the parameters $\nu_{ij}$ (for the case of fading memory) can be fixed as follows. When we deal with the de Sitter model, and $a(t)=a(t_0) e^{H_0 t}$, the term $K_{ij}(t,\xi)$ in (\ref{K1}) takes the form
\begin{equation}
K_{ij}(t,\xi) = K_{ij}^0  H_0 e^{-H_0 \nu_{ij}(t-\xi) }\,.
\label{K17}
\end{equation}
Comparing (\ref{K17}) with (\ref{P3}) we conclude that it is reasonably to assume that the parameters $\nu_{ij}$ are positive.

\section{Key equation of the model}

\subsection{General strategy}

\subsubsection{Balance equations for DE and DM energy densities}

The first step towards resolving the set of equations (\ref{EQ1}), (\ref{EQ2}), (\ref{EQ3}),(\ref{EQ6}), with (\ref{EoS1}) and (\ref{EoS2}) is to obtain the key equation, which contains only one unknown function, namely, the DE energy density scalar $W$. Mention that in the presented model one can consider the functions $W$ and $E$ to depend on cosmological time through the scale factor, i.e., $W = W(a(t))$ and $E = E(a(t))$. This allows us to use the well-known approach based on the introduction of the following dimensionless variable instead of the cosmological time:
\begin{equation}\label{x}
x \equiv \frac{a(t)}{a(t_0)} \,, \quad \frac{d}{dt} = x H(x)\frac{d}{dx}.
\end{equation}
When the Hubble function $H(x)$ is found, the relation between cosmological time and this new variable can be obtained in quadrature as follows:
\begin{equation}\label{quadr}
t-t_0 = \int\limits_1^{\frac{a(t)}{a(t_0)}}\frac{dx}{x H(x)}\,.
\end{equation}
In these terms the balance equations (\ref{EQ2}), (\ref{EQ3}) convert into
\begin{equation}
x\frac{dW}{dx} + 3(W+P) = \omega_{0}(E - W) \,,
\label{Key1}
\end{equation}
\begin{equation}
x\frac{dE}{dx} + 3(E+\Pi) = \omega_{0}(W-E) \,.
\label{Key2}
\end{equation}

\subsubsection{Integral form of the DE and DM equations of state }

In this context, the main interest is connected with the constitutive equations (\ref{EoS1}) and (\ref{EoS2}), which can be now presented in the integral and differential forms. The integral representations are
$$
P(x)=(\Gamma-1)W(x) + x^{-\nu_{11}}K_{11}^0 \int\limits_1^x dy y^{\nu_{11}-1}W(y) +
$$
\begin{equation}
+x^{-\nu_{12}}K_{12}^0 \int\limits_1^x dy y^{\nu_{12}-1}E(y) \,,
\label{Key4}
\end{equation}
$$
\Pi(x)=(\gamma-1)E(x) + x^{-\nu_{21}}K_{21}^0 \int\limits_1^x dy y^{\nu_{21}-1}W(y) +
$$
\begin{equation}
+x^{-\nu_{22}}K_{22}^0 \int\limits_1^x dy y^{\nu_{22}-1}E(y)\,.
\label{Key5}
\end{equation}

\subsubsection{General differential form of the DE and DM equations of state}

It is well known that the integral equations with the multiplicative kernels can be reduced to the differential equations; in our case we obtain the equations of state for the DE (\ref{Key4}) and for the DM (\ref{Key5}) in the following form:
$$
x^2 P^{\prime \prime} + (\nu_{11}+\nu_{12}+1)xP^{\prime} + \nu_{11}\nu_{12}P =
$$
$$
=(\Gamma{-}1)x^2 W^{\prime \prime} + \left[(\Gamma{-}1)(\nu_{11}+\nu_{12}+1) + K_{11}^0\right]xW^{\prime} +
$$
\begin{equation}
+ \left[\nu_{11}\nu_{12}(\Gamma{-}1) {+} K_{11}^0\nu_{12}\right]W {+} K_{12}^0xE' + \nu_{11}K_{12}^0 E \,,
\label{dif1}
\end{equation}
$$
x^2 \Pi^{\prime \prime} + (\nu_{22}+\nu_{21}+1)x\Pi^{\prime} + \nu_{22}\nu_{21}\Pi =
$$
$$
=(\gamma{-}1)x^2 E^{\prime \prime} + \left[(\gamma{-}1)(\nu_{22}+\nu_{21}+1) + K_{22}^0\right]xE^{\prime} +
$$
\begin{equation}
+\left[\nu_{22}\nu_{21}(\gamma{-}1) {+} K_{22}^0\nu_{21}\right]E {+} K_{21}^0xW' {+} \nu_{22}K_{21}^0 W \,.
\label{dif2}
\end{equation}
Here and below the prime denotes the derivative with respect to variable $x$.
Clearly, the equations (\ref{Key1}), (\ref{Key2}), (\ref{Key4}), (\ref{Key5}) do not contain the Hubble function, thus the equation
\begin{equation}\label{Hrel}
3H^2(x)= \Lambda + \kappa \left[W(x)+E(x) \right]
\end{equation}
gives us the unknown function $H(x)$, when $W(x)$ and $E(x)$ are found.

\subsubsection{Initial data problem}

When we convert the integral relationships into the differential equations, we have to keep in mind that the initial data for the quantities participating in these procedure have to satisfy the conditions
$$
P(1)= (\Gamma-1) W(1) \,, \quad \Pi(1)= (\gamma-1) E(1) \,,
$$
$$
P^{\prime}(1) = (\Gamma-1) W^{\prime}(1) + K^0_{11} W(1) + K^0_{12} E(1) \,,
$$
\begin{equation}
\Pi^{\prime}(1) = (\gamma-1) E^{\prime}(1) + K^0_{21} W(1) + K^0_{22} E(1) \,, ... \,.
\label{ini2}
\end{equation}
As well, keeping in mind (\ref{Key1}) and (\ref{Key2}) for the starting point $x=1$ we obtain
\begin{equation}
W^{\prime}(1) = \omega_0 E(1) - W(1)(3\Gamma + \omega_0) \,,
\label{ini22}
\end{equation}
\begin{equation}
E^{\prime}(1) = \omega_0 W(1) - E(1)(3\gamma + \omega_0) \,.
\label{ini24}
\end{equation}
\begin{equation}
W^{\prime}(1)+E^{\prime}(1) =  - 3\left[\Gamma W(1) + \gamma E(1)\right]  \,.
\label{ini27}
\end{equation}
Below we will keep in mind these relationships if we intend to simplify a model and to link some guiding parameters and initial data.

\subsubsection{First particular case: The DM (or DE) self-interaction is absent}

When $K_{22}= 0$ and $\nu_{22}=0$, i.e., when there is no the DM nonlocal self-interaction, the differential version of the equation of state for the dark matter becomes the equation of the first order in derivatives
\begin{equation}
x \Pi^{\prime} {+} \nu_{21}\Pi =(\gamma{-}1) \left(x E^{\prime} {+} \nu_{21} E \right) {+} K_{21}^0 W \,,
\label{dif277}
\end{equation}
linking the state functions $\Pi$, $E$ and $W$. Similarly, when  $K_{11}{=} 0$ and $\nu_{11}{=}0$,  we deal with the differential equation of the first order instead of (\ref{dif1}).

\subsubsection{Second particular case: The DM/DE cross-interaction is absent}

When $K_{12}= 0$ and $\nu_{12}=0$, i.e., when there is no action of the DM on the DE substratum, the corresponding differential version of the equation of state for the dark energy takes the form
\begin{equation}
x P^{\prime} + \nu_{11} P =(\Gamma{-}1)\left(x W^{\prime}+\nu_{11}W \right) + K_{11}^0 W \,.
\label{dif188}
\end{equation}
Again we deal with the equation of the first order in derivatives, but now the DM energy density scalar $E$ disappears from this equation, i.e., the DE constitutive equation can be decoupled from the set of the DE/DM equations of state.
Similarly, when $K_{21}{=} 0$ and $\nu_{21}{=} 0$, the DM constitutive equation happens to be decoupled.

\subsection{The  scheme of reconstruction of the key equation}

\subsubsection{General case: $\omega_0 \neq 0$, $K^0_{ij}\neq 0$, $\nu_{ij} \neq 0$}

When $\omega_0 \neq 0$, we extract the DM energy density $E(x)$ from (\ref{Key1})
\begin{equation}\label{Enrel}
 E(x) = \frac{1}{\omega_0}\left[x W^{\prime}(x) + (3+\omega_0)W + 3P \right] \,,
\end{equation}
and extract the DM pressure $\Pi(x)$ from (\ref{Key2}):
$$
\Pi(x) = - \frac{1}{3\omega_0}\left[x^2 W^{\prime \prime} + (7 + 2\omega_0)xW^{\prime} + (9 + 6\omega_0)W  + \right.
$$
\begin{equation}\label{Pirel}
\left. +3xP^{\prime} + (9 + 3\omega_0)P \right] \,.
\end{equation}
 Then we put $E(x)$ from (\ref{Enrel}) and $\Pi(x)$ from (\ref{Pirel}) to the equations (\ref{dif1}) and (\ref{dif2}), thus excluding the state functions of the dark matter. The last step of this procedure is the following: we exclude the DE pressure $P(x)$ and obtain the key equation for the DE energy density
$$
x^6 W^{(VI)} + \omega_1 x^5 W^{(V)} + \omega_2 x^4 W^{(IV)} + \omega_3 x^3 W^{\prime \prime \prime} +
$$
\begin{equation}
+ \omega_4 x^2 W^{\prime \prime} + \omega_5 x W^{\prime} + \omega_6 W =0  \,.
\label{key100}
\end{equation}
In Appendix I this procedure is described in detail, and the coefficients $\omega_j$ are presented. The main  feature of the key equation (\ref{key100}) is that it is the linear Euler equation of the sixth order in ordinary derivatives, and thus its general solution can be standardly presented in elementary functions. When the solution to the Euler equation (\ref{key100}) is written and $W(x)$ is presented, we find $P(x)$ using (\ref{dif333}); then we find $E(x)$ from (\ref{Enrel}) and $\Pi(x)$ from (\ref{Pirel}). The last step is to find $H(x)$ from (\ref{EQ1}) and then $a(t)$ from (\ref{quadr}).

\subsubsection{The special case $\omega_0 = 0$, but $K^0_{12} \neq 0$ and $K^0_{21} \neq 0$}

The condition $\omega_0 {=}0$ means that the coupled balance equations (\ref{Key1}) and (\ref{Key2}) convert into the conservation laws for the DE and DM individually, however, there exists the nonlocal cross-interaction between DE and DM. When the parameter $\omega_0$ vanishes, we have to change the strategy of the derivation of the key equation. Now we extract the DE pressure $P$ from (\ref{Key1}) and the DM pressure $\Pi$ from (\ref{Key2})
\begin{equation}
P = - \frac13 x W^{\prime} - W \,,
\quad
\Pi = - \frac13 x E^{\prime} - E \,.
\label{Key29}
\end{equation}
Then we put these $P$ and $\Pi$ into the equations (\ref{dif1}) and (\ref{dif2}) obtaining two equations, which link now the DE and DM energy densities $W$ and $E$. At the last step we exclude $E(x)$ and obtain the key equation for the DE energy density of the form (\ref{key100}), but all the coefficients $\omega_j$ should be replaced by $\Omega_j$ (they are presented in the Appendix II).
We deal again with the Euler equation of the sixth order in derivatives; when $W(x)$ is written, we obtain $E(x)$ from (\ref{EW0}), then find $P(x)$ and $\Pi(x)$ from (\ref{Key29}), $H(x)$ from (\ref{EQ1}) and $a(t)$ from (\ref{quadr}).

\subsubsection{The special case $\omega_0 =0$ and $K^0_{12}=0$ }

As an example, we discuss now the case, when the DM action on the dark energy is assumed to be negligible. Now we obtain that the equation for the DE energy density $W$ happens to be decoupled
\begin{equation}
x^3 W^{\prime \prime \prime} {+} \alpha_1 x^2 W^{\prime \prime}  {+}
 \alpha_2 x W^{\prime}  {+} \alpha_3 W = 0 \,,
\label{Key517}
\end{equation}
where the coefficients $\alpha_1$, $\alpha_2$, $\alpha_3$ are presented in Appendix II.
We deal now with the Euler equation of the third order for the unknown function $W$. When $W(x)$ is found, we solve the Euler equation of the third order (\ref{Key52}) for $E(x)$ and then follow the logics of the previous subsubsection.

\subsubsection{The special case $\omega_0 \neq 0$, but $K^0_{12}=0$ and $\nu_{12}=0$}

Now we find that the coefficient $\omega_6$ vanishes, $\omega_6{=}0$ (see Appendix I); this means that the key equation becomes of the fifth order in derivatives. The same result appears if $K^0_{21}{=}0$ and $\nu_{21}{=}0$.
When the pairs of the coefficients $K^0_{ij}$ and the corresponding pairs of $\nu_{ij}$ vanish, one can reduce the key equation (\ref{key100}) to the differential equation of the fourth order.

\section{Analysis of the solutions to the key equation}

\subsection{Characteristic equation and the structure of the general solution}

The general solution to the Euler equation (\ref{key100}) can be reconstructed by the standard method: we search for the particular solutions in the form $W(x) \to x^{\sigma}$, and obtain the characteristic equation of the sixth order for $\sigma$:
\begin{equation} \label{characteristic}
\sigma^6 + (\omega_1-15)\sigma^5 + \left(85-10\omega_1 + \omega_2\right)\sigma^4 +
\end{equation}
$$
+  \sigma^3 (-225 + 35 \omega_1 -6 \omega_2 +\omega_3) +
$$
$$
+\left(274-50\omega_1 +11\omega_2 -3 \omega_3 + \omega_4 \right) \sigma^2 +
$$
$$
+\left({-}120 {+}24 \omega_1 {-} 6\omega_2 {+} 2\omega_3 {-}\omega_4 {+} \omega_5\right)\sigma {+}  \omega_6=0 \,.
$$
Clearly, the algebraic polynomial of the sixth order with real coefficients can have the following sets of roots: six real roots; four real roots and a pair of complex conjugated ones; two real roots and two pairs of complex conjugated ones; three pairs of complex conjugated roots.  The general solution to the Euler equation (\ref{key100}) is known to be the linear combination of six fundamental solutions. For the simple real root $\sigma_1$ the basic solution is $x^{\sigma_1}$; when $k$ real roots coincide the corresponding basic solutions are $x^{\sigma_1}$, $ x^{\sigma_1} \log{x}$, ..., $x^{\sigma_1} (\log{x})^{k-1}$. When there is a complex conjugated pair $\sigma=\alpha \pm i \beta$ among the roots of the characteristic equation, one has to choose two basic solutions in the form $x^{\sigma_1}\cos{(\beta \log{x})}$ and $x^{\sigma_1}\sin{(\beta \log{x})}$; when there are coinciding pairs of the complex conjugated roots, one has to use the products of the corresponding basic functions with $(\log{x})^s$, as in the case of the real roots. This procedure is standard, and below we present the classification of the mentioned six roots and consider the corresponding solutions to the key equation (\ref{key100}) in the asymptotic regime $x \to \infty$ in order to select the appropriate models, which seem to be physically motivated.

\subsection{Real roots}
Let us start with the case when all six roots of the characteristic equation (\ref{characteristic}), $\sigma_{(a)}$ are real.
One can distinguish eleven different subcases: there is one the so-called completely non-degenerated set of real roots (there are no coinciding pairs); also there are ten degenerated sets (two, three, four, five or six roots coincide, etc.).

\subsubsection{Completely non-degenerate set of roots}

When all six roots do not coincide, the general solution to the key equation (\ref{key100}) can be written as follows:
\begin{equation}\label{W1}
	W(x) = \sum\limits_{(a)=1}^{6}C_{(a)} x^{\sigma_{(a)}} \,,
\end{equation}
where $C_{(a)}$ are integration constants. According to the scheme of analysis proposed above we obtain now that the solutions for $E(x)$, $P(x)$, $\Pi(x)$ and $H^2$ have exactly the same structure as $W(x)$ (\ref{W1}), we just have to specify the corresponding coefficients.
Also, the acceleration parameter
\begin{equation}\label{q1}
-q \equiv \frac{\ddot{a}}{aH^2} = 1{+} \frac{x(H^2(x))^{\prime}}{2H^2(x)} =
\end{equation}
$$
=\frac{1{+} \frac{3}{\Lambda}\sum\limits_{(a)=1}^{6}\bar{C}_{(a)}\left(1+\frac12 \sigma_{(a)} \right) x^{\sigma_{(a)}}}{1{+} \frac{3}{\Lambda}\sum\limits_{(a)=1}^{6}\bar{C}_{(a)} x^{\sigma_{(a)}}}
$$
happens to be presented in the elementary functions.

The scale factor $a(t)$ is the function, which generally can be found in quadratures only
\begin{equation}\label{a1}
	t-t_0 = \int\limits_1^{\frac{a(t)}{a(t_0)}}\frac{dx}{x \sqrt{\frac{\Lambda}{3} + \sum\limits_{(a)=1}^{6}\bar{C}_{(a)} x^{\sigma_{(a)}}}} \,.
\end{equation}
Clearly,  in general the scale factor can be obtained only numerically. However, one can analyze the asymptotic behavior of the Universe geometric characteristics as follows.
Let the root  of the characteristic equation $\sigma_{(\rm m)}$ be the maximal among the six real roots $\sigma_{(a)}$.  The root $\sigma_{(\rm m)}$ can be positive, zero or negative.

\noindent
1. $\sigma_{(\rm m)}>0$.

\noindent
In this case the scale factor behaves asymptotically as
\begin{equation}\label{a2}
x= \frac{a(t)}{a(t_0)} \propto \left[\frac{1}{(t_{*}-t)}\right]^{\frac{2}{\sigma_{(m)}}}  \,.
\end{equation}
When $t \to t_*$, the functions $W(t)$, $E(t)$, $P(t)$, $\Pi(t)$ tend to infinity, and we deal with the Big Rip (for the classification of future singularities see, e.g., \cite{Rip1,Rip2,Rip3,Rip4}).

\noindent
2. $\sigma_{(\rm m)}=0$.

\noindent
Now we deal with the behavior, which is characterized by the asymptotically constant DE and DM state functions: $W(t) \to W_{\infty}$, $E(t) \to E_{\infty}$, $P(t) \to P_{\infty}$, $\Pi(t) \to \Pi_{\infty}$ and $H(t)\to H_{\infty}$; as for the scale factor, it behaves as
\begin{equation}\label{a3}
a(t) \propto e^{H_{\infty}t}  \,, \quad H_{\infty} = \sqrt{\frac{\Lambda + \kappa (W_{\infty} + E_{\infty})}{3}} \,.
\end{equation}
We follow the works \cite{Rip1,Rip2,Rip3,Rip4} and prefer to indicate this final state as Pseudo Rip.

\noindent
3. $\sigma_{(\rm m)}<0$.

\noindent
For this case all the DE and DM state functions tend asymptotically to zero, and we deal with the standard de Sitter asymptote with $H_{\infty} = \sqrt{\frac{\Lambda}{3}}$.

\begin{figure}[H]
	\centering
	\includegraphics[width=\columnwidth]{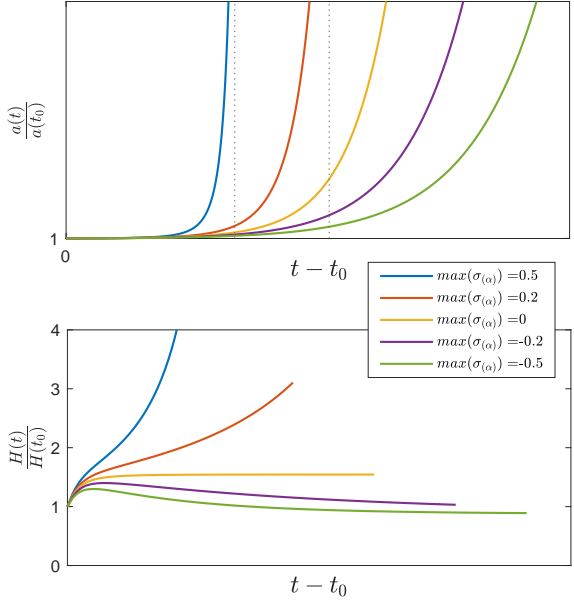}
	\caption{This figure illustrates the behavior of dimensionless scale factor $a(t)/a(t_0)$ (upper panel) and Hubble function $H(t)/H(t_0)$ (bottom panel) as functions of cosmological time $t-t_0$;
 five values of the $max(\sigma_{(\alpha)})$ are fixed in the right side of the figure. $C_{(m)}$ corresponding to the maximal root $\sigma_m$ is considered to be positive;
 other integration constants $C_{(i)}$ are negative.}
	\label{fig1}
\end{figure}

Illustration of the mentioned regimes is presented in Fig.\ref{fig1}.

\subsubsection{Degenerated sets of roots}

When all the roots of the characteristic equations are real, there exists ten specific cases, which describe the situations with coinciding roots:

\noindent
1) two roots coincide, say, $\sigma_{(1)}{=}\sigma_{(2)}$, and other roots are different;

\noindent
2) three roots coincide, say, $\sigma_{(1)}{=}\sigma_{(2)}{=}\sigma_{(3)}$, and other roots are different;

\noindent
3) four roots coincide, say, $\sigma_{(1)}{=}\sigma_{(2)}{=}\sigma_{(3)} {=} \sigma_{(4)}$, and other roots are different;

\noindent
4) five roots coincide, say, $\sigma_{(1)}{=}\sigma_{(2)}{=}\sigma_{(3)} {=} \sigma_{(4)}{=} \sigma_{(5)}$, and the last one differs from them;

\noindent
5) all six roots coincide;

\noindent
6) there are two pairs of coinciding roots, say, $\sigma_{(1)}{=}\sigma_{(2)} \neq \sigma_{(3)} {=} \sigma_{(4)}$ and other two roots are different;

\noindent
7) there are three pairs of coinciding roots, say, $\sigma_{(1)}{=}\sigma_{(2)} \neq \sigma_{(3)} {=} \sigma_{(4)} \neq \sigma_{(5)} {=} \sigma_{(6)}$;

\noindent
8) there is the set of roots, satisfying the conditions $\sigma_{(1)}{=}\sigma_{(2)} {=} \sigma_{(3)} \neq \sigma_{(4)} {=} \sigma_{(5)} \neq \sigma_{(6)}$;

\noindent
9) there is the set of roots, satisfying the conditions $\sigma_{(1)}{=}\sigma_{(2)} {=} \sigma_{(3)} {=} \sigma_{(4)} \neq  \sigma_{(5)} {=} \sigma_{(6)}$;

\noindent
10) there are two trio of coinciding roots, $\sigma_{(1)}{=}\sigma_{(2)} {=} \sigma_{(3)} \neq \sigma_{(4)} {=} \sigma_{(5)} {=} \sigma_{(6)}$;

\noindent
The procedure of representation of the general solution to the key equation in all ten cases is well documented.
For instance, let $k$ real roots coincide ($2 \leq k \leq 6$), say,
$ \sigma_{(1)} {=} \sigma_{(2)} {=} ...{=} \sigma_{(k)} \equiv \sigma_0$, and other roots be different. The corresponding general solution to the key equation reads	
\begin{equation}\nonumber
W(x) = 	x^{\sigma_{0}}\left[\tilde{C}_{1} {+} \tilde{C}_{2} \log{x} {+} ... {+} \tilde{C}_{k} \log^{k-1}{x} \right]  {+}
\end{equation}
\begin{equation}\label{kkk}
	+ \sum_{j{=}k{+}1}^{(6)} \tilde{C}_{j} x^{\sigma_{(j)}} \,.
\end{equation}

This structure is also typical for the functions $E(x)$, $P(x)$, $\Pi(x)$ and $H^2(x)$.

The asymptotic behavior of the solutions can be estimated keeping in mind two principal cases: what is bigger: $\sigma_0$ or one of the roots $\sigma_{(j)}$, $j>k$. When $\sigma_{(j)}>\sigma_0$, the asymptotic behavior of the system is described in the previous subsubsection. Now we assume that the root $\sigma_0$ is the biggest one among the roots of the set under discussion. Then the biggest term in (\ref{kkk}) is
\begin{equation}\label{kkk2}
W(x \to \infty) \to  \tilde{C}_{k} \ x^{\sigma_{0}}  (\log{x})^{k-1} \,.
\end{equation}
The solution for the scale factor $a(t)$ depends on the sign of the root $\sigma_0$.

\noindent
1. When $\sigma_0<0$, $H(x) \to \sqrt{\frac{\Lambda}{3}}$ and we deal with the de Sitter type behavior of the model.

\noindent
2. When $\sigma_0=0$, and thus $H(x) \to {\cal K} \left(\log{x} \right)^{\frac{k{-}1}{2}}$,  we obtain three interesting cases.

\noindent
2.1.  If $k=2$, $a(t \to \infty) \propto e^{\frac14 {\cal K}^2 t^2}$; we deal with the solution indicated as anti-Gaussian solution in \cite{Arc1}. It is an example of the Little Rip type behavior.

\noindent
2.2. If $k=3$, $a(t \to \infty) \propto e^{e^{{\cal K} t}}$; we deal with the solution indicated as super-exponential solution in \cite{Arc1}. Again,  it is an example of the Little Rip behavior.

\noindent
2.3. If $3<k\leq 6$,  the integral (\ref{quadr}) converges when the upper limit tends to infinity. This means that the scale factor $a$ reaches the infinite value during the finite interval of the cosmological time. We deal now with the  Big Rip type solution

\noindent
3.  When $\sigma_0>0$, and thus $H(x) \to {\cal K} x^{\frac12 \sigma_0} \left(\log{x} \right)^{\frac{k{-}1}{2}}$, we can rewrite (\ref{a1}) as follows
\begin{equation}\label{kk199}
\left(\frac{\sigma_0}{2} \right)^{\frac{3{-}k}{2}} {\cal K} t \to \gamma(s,u) \equiv \int\limits_0^u d\xi e^{-\xi} \xi^{s-1} \,,
\end{equation}
where $\gamma(s,u)$ is the incomplete lower Gamma function with the arguments
\begin{equation}\label{kk299}
s= \frac{3{-}k}{2} \,, \quad u= \frac{\sigma_0}{2} \log{\left[\frac{a(t)}{a(t_0)}\right]} \,.
\end{equation}
This function links with the complete Gamma function $\Gamma(s)$ and with incomplete upper gamma function
$\Gamma(s,u)= \int\limits_u^{\infty} d\xi e^{-\xi} \xi^{s{-}1}$
by the simple condition $\gamma(s,u) = \Gamma(s) {-} \Gamma(s,u)$. For large argument $u$ we can use the relationship
$\gamma(s,u) \approx \Gamma(s) {-} e^{{-}u} u^{s{-}1} $. In other words, the integral (\ref{kk199}) converges at $a \to \infty$, and the scale factor reaches the infinite value during the finite interval of time; again we deal with the Big Rip scenario.

\vspace{3mm}

\noindent
{\it SHORT RESUME}

We assume that the Big Rip scenario are not physically motivated, thus, the models with real characteristic roots have to correspond to two cases: first, $\sigma_{(a)} < 0$, second, $\sigma_{(a)} = 0$ and $k \leq 3$.

\subsection{Complex roots}

\subsubsection{Preliminary classification}

When not all the roots are real, we obtain six intrinsic cases.

\noindent
1. There is one pair of complex conjugated roots
$\sigma_{(a)} = \alpha_{(a)} \pm i \beta_{(a)} $, and other four roots are real (there are five internal cases with and without degeneracy of the real roots).

\noindent
2. There are two different pairs of complex conjugated roots, and two roots are real (there are two internal cases).

\noindent
3. There are three different pairs of complex conjugated roots.

\noindent
4. There are two coinciding pairs of complex conjugated roots, and two roots are real (there are two internal cases).

\noindent
5. There are three pairs of complex conjugated roots, and two of them coincide.

\noindent
6. There are three coinciding pairs of complex conjugated roots.

\subsubsection{First case: there is one pair of complex conjugated roots }

Now the solution to the key equation (\ref{key100}) can be written as follows:
\begin{equation}
W(x) = x^{\alpha} \left[C_1 \cos\left(\beta \log{x} \right) {+} C_{2} \sin\left(\beta \log{x}\right) \right]+ W_{(\rm real)} \,,
\end{equation}
where the decomposition $W_{(\rm real)}$ is given in the previous subsection. Again we see that the DE and DM state functions $E(x)$, $P(x)$, $\Pi(x)$, as well as, the square of the Hubble function $H^2$ have the same form, but the coefficients of decomposition are specific.
A principally new details of solution appear, when the real part $\alpha$ of the complex root happens to be bigger than the real roots $\sigma_{(a)}$ encoded in the term $W_{(\rm real)}$. Then in the asymptotic regime $x \to \infty$ we obtain the following integral for searching for the scale factor:
\begin{equation}\label{complex7}
t{-}t_0  {=} \int\limits^{\frac{a(t)}{a(t_0)}}_{1} \frac{d \log{x}}{\sqrt{\frac{\Lambda}{3} {+} x^{\alpha} \left[\tilde{C}_1 \cos{(\beta \log{x})} {+} \tilde{C}_{2} \sin(\beta \log{x})\right]}} \,.
\end{equation}
For illustration, we consider $\alpha=0$ and $\tilde{C}_{2}=0$, $\tilde{C}_{1} =C$, obtaining
\begin{equation}\label{complex27}
t-t_0 = \int\limits^{\frac{a(t)}{a(t_0)}}_{1} \frac{d \log{x}}{\sqrt{\frac{\Lambda}{3} + C \cos{(\beta\log{x})}}} \,.
\end{equation}
Our goal is to obtain the formula
\begin{equation}\label{elliptic}
\beta k \sqrt{\frac{C}{2}}(t-t_0) = \int\limits_{0}^{\varphi} \frac{d\theta}{\sqrt{1-k^2\sin^2\theta}} \,,
\end{equation}
which appears from (\ref{complex27}) after the following redefinitions:
\begin{equation}\label{elliptic1}
\theta \equiv \frac12 \beta \log{x} \,, \quad \varphi = \frac{1}{2}\beta\log\left(\frac{a(t)}{a(t_0)}\right) \,, \quad k^2 = \frac{6C}{\Lambda{+}3C} \,.
\end{equation}
This interest is predetermined by the fact that the right-hand side of (\ref{elliptic}) presents the definition of the incomplete elliptic integral of the first kind $F\left(\varphi \ | \ k^2\right)$, and that the functions reciprocal to $u=F\left(\varphi \ | \ k^2\right)$ are connected with the Jacobi elliptic sine and cosine functions
\begin{equation}\label{elliptic2}
\textrm{sn}(u,k^2)=\sin \varphi \,, \quad \textrm{cn}(u,k^2)=\cos \varphi \,,
\end{equation}
and two auxiliary functions
\begin{equation}\label{elliptic3}
\textrm{dn}(u,k^2)=\sqrt{1-k^2\sin^2 \varphi} \,, \quad \textrm{am}(u,k^2)= \varphi \,.
\end{equation}
The abbreviation $\textrm{am}$ is used for the so-called Jacobi amplitude function.

Now we see that the cosmological time can be expressed via the incomplete elliptic integral
\begin{equation}\label{elliptic4}
t-t_0 = \frac{2\sqrt{3}}{\beta\sqrt{\Lambda + 3C}} \  F\left(\frac{1}{2}\beta\log{x}\ {\bigg|} \ \frac{6C}{\Lambda + 3C}\right)\,,
\end{equation}
and the scale factor is of the form
\begin{equation}\label{elliptic5}
\frac{a(t)}{a(t_0)} =  \exp{\left[\frac{2}{\beta} am \left(\frac{\beta\sqrt{\Lambda+3C}}{2\sqrt{3}}(t{-}t_0), \ \frac{6C}{\Lambda{+}3C}\right)\right]} \,.
\end{equation}
There are two special regimes of behavior of the presented solution, when $k^2=1$ and $k^2=0$.

\vspace{3mm}
\noindent
{\it (i)} When $C=\frac{\Lambda}{3}$ and thus $k^2=1$, the elliptic functions are known to be converted into the hyperbolic functions
\begin{equation}\nonumber
\sin{\varphi} = \textrm{sn}(u,1) = \tanh{u} \,, \quad \cos \varphi = \frac{1}{\cosh{u}} \,,
\end{equation}
\begin{equation}\label{elliptic6}
	\quad \textrm{am}(u,1) = \varphi = \arcsin{(\tanh{u})}\,.
\end{equation}

In this case (\ref{elliptic5}) gives
\begin{equation}\label{elliptic55}
\frac{a(t)}{a(t_0)} =  \exp\left\{\frac{2}{\beta} \arcsin{\left[\tanh{\left(\beta \sqrt{\frac{\Lambda}{6}}(t{-}t_0) \right)}\right]} \right\} \,.
\end{equation}
In the asymptotic regime, when $t \to \infty$, we obtain that the scale factor tends to the constant value $a(t_0)e^\frac{\pi}{\beta}$, and the Hubble function tends to zero,
$H \to 0$.

\vspace{3mm}
\noindent
{\it (ii)} When $C=0$ and thus $k^2=0$, we obtain from (\ref{complex27}) that $a(t)=a(t_0)e^{\sqrt{\frac{\Lambda}{3}}(t-t_0)}$, i.e., we deal with the de Sitter regime.

When the parameter $k^2$ belongs to the interval $0<k^2<1$, i.e., $0<C<\frac{\Lambda}{3}$, we see that $t \to \infty$ at  $a(t) \to \infty$; as for the Hubble function $H(t)$, it remains bounded,
$\sqrt{\frac{\Lambda}{3}-C} < H < \sqrt{\frac{\Lambda}{3}+C}$.

\begin{figure}[H]
	\centering
	\includegraphics[width=\columnwidth]{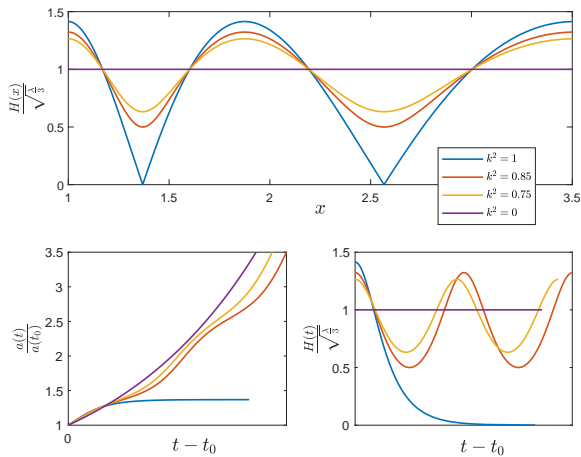}
	\caption{Upper panel: Illustration of the behavior of the Hubble function $\frac{H(x)}{\sqrt{\Lambda/3}}$. Here $x = \frac{a(t)}{a(t_0)}$ is the dimensionless scale factor, $\Lambda$ is cosmological constant, $t$ is cosmological time.
	Bottom panels: Dimensionless scale factor $a(t)/a(t_0)$ (left panel) and Hubble function $\frac{H(t)}{\sqrt{\Lambda/3}}$ (right panel). Blue line corresponds to the case $k^2=1$ or  $C = \frac{\Lambda}{3}$; red line to $C = \frac{\Lambda}{4}$; orange line to $C = \frac{\Lambda}{5}$; purple line to $C=0$.}
	\label{fig2}
\end{figure}

The regime is illustrated in Fig.\ref{fig2}.

\subsubsection{Second case: two complex conjugated pairs coincide}

In order to illustrate the novelty, which appears in the asymptotic behavior of the system in this case, we consider the model, in which the real parts of two coinciding pairs of roots are equal to zero $\sigma_{1}=\sigma_{2} = \pm i\beta$, and the real parts of all other roots are non-positive $Re  \sigma_{(a)} \leq 0$.
Then in the asymptotic regime $x \to \infty$
\begin{equation}\label{21}
	W(x) \to \log{x} \left(C_1 \cos\beta\log{x} {+} C_{2}\sin\beta\log{x}\right) \,.
\end{equation}
The state functions $E(x)$, $P(x)$, $\Pi(x)$, as well as, the square of the Hubble function $H^2(x)$ have the same structure.
Searching for the scale factor for the late-time Universe evolution we have to calculate the following integral:
\begin{equation}\label{complexintegral2}
	t-t_0 = 2 \int\limits^{\frac{a(t)}{a(t_0)}}_{1} \frac{d \sqrt{\log{x}}}{\sqrt{C_1 \cos{(\beta\log{x})} {+} C_{2}\sin{(\beta\log{x})}}} \,.
\end{equation}
Clearly, the term in the square root takes zero value at $x=x_*$, where $\tan{(\beta \log{x_*})}=-\frac{C_1}{C_2}$, and then changes the sign. This means that the Hubble function becomes imaginary and the model happens to be inappropriate.

\subsubsection{Third case: three complex conjugated pairs coincide}

When three complex conjugated pairs coincide and have the form $\alpha \pm i \beta$,
the square of the Hubble function at $x \to \infty$ can be approximated as
\begin{equation}\label{qq}
H^2(x) \to  x^{\alpha} \log^2{x} \left[\tilde{C}_{1} \cos{(\beta\log{x})} {+} \tilde{C}_{2}\sin{(\beta\log{x})}\right].
\end{equation}
Again, there exists a value of the scale factor $x_*$, when the Hubble function takes zero value and then becomes the imaginary one; this model is not appropriate.

\vspace{3mm}
\noindent
{\it SHORT RESUME}

We assume that physically motivated models with complex conjugated pairs of the characteristic roots correspond to the case, when there are no coinciding pairs.

\section{First example of exactly integrable models: Pressureless dark matter and non-locally self-interacting dark energy are coupled by the local link }

\subsection{Truncated model}

In this first model we assume that  $\gamma=1$ and $K^0_{21}=K^0_{22}=0$, $\nu_{21}=\nu_{22}=0$. In this case according to (\ref{Key5})  the dark matter is pressureless, i.e., $\Pi(x)=0$.
The interaction with the DE is considered to be local, i.e., $\omega_0 \neq 0$, and the dark energy to be characterized by $\Gamma =0$ and $K^0_{12}=0$, $\nu_{12}=0$, i.e., the DE self-interaction is non-local.
The scheme of derivation of the key equation is now simplified; we obtain the DE pressure and the DM energy density in the form
\begin{equation}
P(x) = \frac{x^2 W^{\prime \prime} {+} 2x W^{\prime}(2 {+} \omega_0) {+} 3W(3{+}2\omega_0 {-} \nu_{11} {+} K^0_{11})}{3 \left(\nu_{11} {-}3 {-} \omega_0 \right)}\,,
\label{qq10}
\end{equation}
\begin{equation}
E(x){=} \frac{x^2 W^{\prime \prime} {+} x W^{\prime}(1 {+} \omega_0 {+} \nu_{11}) {+} W( 3K^0_{11} {+} \omega_0 \nu_{11} {-} \omega^2_0)}{ \omega_0 \left(\nu_{11} {-}3 {-} \omega_0\right)},
\label{qq9}
\end{equation}
where the DE energy density $W(x)$ satisfies the Euler equation of the third order
\begin{equation}\nonumber
x^3 W^{\prime \prime \prime} + x^2 W^{\prime \prime} \left(6+ 2\omega_0 + \nu_{11} \right) +
\end{equation}
\begin{equation}\nonumber
+x W^{\prime}\left(4+ 5\omega_0 + 4 \nu_{11} + 3K^0_{11} + 2 \nu_{11} \omega_0 \right) +
\end{equation}
\begin{equation}\label{qq11}
+ 3W \left[\nu_{11} \omega_0 + K^0_{11}(3+\omega_0) \right] =0 \,.
\end{equation}
If we calculate the third derivative of this equation we obtain (\ref{key100}) with $\omega_6=\omega_5=\omega_4=0$.

The corresponding characteristic equation of the third order
$$
\sigma^3 {+} \sigma^2 (3{+}2\omega_0 {+} \nu_{11}) {+} \sigma (3 \omega_0 {+} 3\nu_{11} {+} 3K^0_{11} {+} 2 \nu_{11} \omega_0) +
$$
\begin{equation}
 + 3[K^0_{11}(3+ \omega_0)+ \nu_{11}\omega_0]  = 0
\label{qq1}
\end{equation}
can have three real root or a pair of complex conjugated roots plus one real root. In order to simplify the illustration of general scheme of the solution classification, we consider the following choice of the parameter $K^0_{11}$:
\begin{equation}
K^0_{11} = - \frac{\nu_{11}\omega_0}{3+ \omega_0} \,,
\label{qq2}
\end{equation}
thus providing that the first root, $\sigma_1=0$, is real and the roots $\sigma_2$, $\sigma_3$ satisfy the quadratic equation
\begin{equation}
\sigma^2 {+} \sigma (3{+}2\omega_0 {+} \nu_{11}) {+} \left[3 (\omega_0 {+} \nu_{11}) {+} \frac{(3+2\omega_0) \nu_{11}\omega_0}{3+\omega_0}\right] =0 \,.
\label{qq3}
\end{equation}

\subsection{Three coinciding real roots}

We start the illustration with the model, which admits three coinciding roots; now they are $\sigma_1=\sigma_2=\sigma_3=0$. It is possible, when
\begin{equation}
\nu_{11}=-(3+2\omega_0) \,, \quad K^0_{11} = \frac{\omega_0 (3+2\omega_0)}{3+\omega_0} \,,
\label{qq32}
\end{equation}
and the parameter $\omega_0$ is the solution to the equation
\begin{equation}
\omega_0 (3+2\omega_0)^2 +3 (3+\omega_0)^2 = 0 \,.
\label{qq33}
\end{equation}
The equation (\ref{qq33}) has only one real root $\omega_0 \approx -2.06$.
The solution to the key equation is
\begin{equation}
W(x)= W(1)+ C_2 \log{x} + C_3 \log^2{x} \,,
\label{qq34}
\end{equation}
where $C_2$ and $C_3$ are the integration constants, which can be found from the initial conditions as follows:
\begin{equation}
C_2 {=} W^{\prime}(1) , \quad C_3 {=} \frac{3\omega_0 (2{+}\omega_0)}{2(3{+}\omega_0)^2} \left[E^{\prime}(1)(3{+}\omega_0) {-} W^{\prime}(1) \omega_0 \right].
\label{qq53}
\end{equation}
The DE pressure $P(x)$ and the DM energy density can be presented as follows
\begin{equation}\nonumber
P(x)= P(1) -
C_3 \frac{(3{+}2\omega_0)}{(3{+}\omega_0)} \log^2{x} -
\end{equation}
\begin{equation}\label{qq35}
- \frac{(3{+}2\omega_0)[2C_3(3{+}\omega_0){+} 9C_2(2{+}\omega_0)]}{9(2{+}\omega_0)(3{+}\omega_0)} \log{x} \,,
\end{equation}
\begin{equation}\nonumber
E(x)= E(1) +
C_3 \frac{\omega_0}{(3{+}\omega_0)} \log^2{x} +
\end{equation}
\begin{equation}\label{qq36}
+ \frac{[2C_3(3{+}\omega_0)^2{+} 3C_2(2{+}\omega_0)\omega_0^2]}{3\omega_0(2{+}\omega_0)(3{+}\omega_0)} \log{x} \,.
\end{equation}
The Hubble function $H(x)$ can be written as
\begin{equation}
H(x)= \pm \sqrt{H^2(1)  +  h_2 \log^2{x} + h_3 \log{x} } \,,
\label{qq38}
\end{equation}
where the following guiding parameters are introduced
\begin{equation}
H^2(1) = \frac{\Lambda}{3} {+} \frac{\kappa}{3} [W(1){+}E(1)] \,,
\label{qq41}
\end{equation}
\begin{equation}
h_2 {=}  \frac{\kappa \omega_0 (3{+}2\omega_0) (2{+}\omega_0)}{2(3{+}\omega_0)^2} \left[E^{\prime}(1) {-}  \frac{\omega_0}{(3{+}\omega_0)} W^{\prime}(1) \right],
\label{qq42}
\end{equation}
\begin{equation}
h_3 = \frac{\kappa}{3} \left[E^{\prime}(1){+} W^{\prime}(1) \right] \,.
\label{qq43}
\end{equation}
The further results depend essentially on the sign of the parameter $h_2$, or equivalently, on the relationships between initial values of the derivatives of the DE and DM energy densities.

\subsubsection{$h_2 >0$, Little Rip models}

This is possible, when $E^{\prime}(1) <  -  \frac{|\omega_0|}{(3-|\omega_0|)} W^{\prime}(1)$.
The result depends now on the relationship between $W(1)$, $E(1)$, $W^{\prime}(1)$, $E^{\prime}(1)$ and $\Lambda$, but for the sake of compactness we formulate the corresponding conditions using the parameters $H(1)$, $h_2$, $h_3$ and their combinations.

\vspace{3mm}

\noindent
{\it (i)} The first case $H^2(1)> \frac{h^2_3}{4h_2}$.

\noindent
We obtain for the scale factor the following formula:
\begin{equation}
a(t) = a(t_{\times}) \exp\left\{h_{\times} \sinh{[\sqrt{h_2}(t{-}t_{\times})]} \right\}
\,,
\label{LR1}
\end{equation}
where the auxiliary quantities are
\begin{equation}
a(t_{\times}) = a(t_0) \exp\left({-} \frac{h_3}{2h_2} \right) \,,
\label{LR2}
\end{equation}
\begin{equation}
h_{\times} = \frac{1}{\sqrt{h_2}} \sqrt{H^2(1){-} \frac{h^2_3}{4h_2}} \,,
\label{LR21}
\end{equation}
\begin{equation}
t_{\times} = t_0 - \frac{1}{\sqrt{h_2}} {\rm Arsh}\left[\frac{h_3}{2\sqrt{h_2} \sqrt{H^2(1)- \frac{h^2_3}{4h_2}}} \right] < t_0 \,.
\label{LR3}
\end{equation}
The presented solution for the scale factor is regular; its asymptotic behavior can be indicated as super-exponential.
In terms of the cosmological time the Hubble function can be presented as
\begin{equation}
H(t) = \sqrt{H^2(1){-} \frac{h^2_3}{4h_2}} \cosh{[\sqrt{h_2}(t{-}t_{\times})]} \,,
\label{LR5}
\end{equation}
and the acceleration parameter is
\begin{equation}
{-}q = 1{+} \frac{\dot{H}}{H^2} = 1{+} \frac{1}{h_{\times}}\left\{\frac{\sinh{[\sqrt{h_2}(t{-}t_{\times})]}}{\cosh^2{[\sqrt{h_2}(t{-}t_{\times})]}} \right\}.
\label{LR6}
\end{equation}
Asymptotically, $H(t \to \infty) \to \infty$, and $-q(t \to \infty) \to 1$.

\vspace{3mm}

\noindent
{\it (ii)} The second case $H^2(1)< \frac{h^2_3}{4h_2}$.

\noindent
Formally speaking, now, in order to obtain the scale factor we have to replace the function $\sinh$ with $\cosh$, and  $\sqrt{H^2(1){-} \frac{h^2_3}{4h_2}}$ with $\sqrt{\frac{h^2_3}{4h_2}{-}H^2(1)}$. We do not discuss the details of this exact solutions, since they are similar to the previous case.

\vspace{3mm}

\noindent
{\it (iii)} The third case $H^2(1)= \frac{h^2_3}{4h_2}$.

\noindent
We obtain the following solutions for the scale factor, Hubble function and acceleration parameter, respectively:
\begin{equation}
a(t) = a(t_0) \exp\left\{ \frac{h_3}{2h_2}\left[e^{\sqrt{h_2}(t-t_0)} -1 \right] \right\} \,,
\label{LR7}
\end{equation}
\begin{equation}
H(t) = H(t_0) e^{\sqrt{h_2}(t-t_0)} \,,
\label{LR8}
\end{equation}
\begin{equation}
-q = 1+ \frac{\dot{H}}{H^2} = 1+ \frac{\sqrt{h_2}}{H(t_0)} e^{- \sqrt{h_2}(t-t_0)}\,.
\label{LR9}
\end{equation}

\vspace{3mm}
\noindent
{\it SHORT RESUME }

\noindent
Three submodels discussed above are regular, and are characterized by the super-exponential asymptotes for the scale factor, exponential asymptotes for the Hubble function, DM and DE energy density scalars and DE pressure. We deal with variants of the Little Rip, for which the infinite values of the state functions can be reached during infinite time interval.

\subsubsection{$h_2<0$, quasi-periodic models}

Such a situation can be realized, when $E^{\prime}(1)> -  \frac{|\omega_0|}{(3-|\omega_0|)} W^{\prime}(1)$. In this situation we obtain the exact solution for the scale factor of the periodic type
\begin{equation}
a(t) = a(t_{+}) \exp{\left\{ h_{+} \sin{[\sqrt{|h_2|}(t{-}t_{+})]} \right\}} \,,
\label{P1}
\end{equation}
\begin{equation}
a(t_{+})= a(t_0) \exp\left(-\frac{h_3}{2|h_2|}\right) \,,
\label{P2}
\end{equation}
\begin{equation}
h_{+} = \frac{1}{\sqrt{|h_2|}} \sqrt{H^2(1){+} \frac{h^2_3}{4|h_2|}} \,,
\label{P21}
\end{equation}
\begin{equation}
t_{+} {=} t_0 {-} \frac{1}{\sqrt{|h_2|}} {\rm arcsin}\left[\frac{h_3}{2\sqrt{|h_2|} \sqrt{H^2(1){+} \frac{h^2_3}{4|h_2|}}} \right] \,.
\label{P37}
\end{equation}
The parameter $\sqrt{|h_2|}$ plays the role of the frequency of the oscillations, and the parameter $a(t_{+})$ describes the mean value of the Universe radius. The Hubble function
\begin{equation}
H(t) = \sqrt{H^2(1){+} \frac{h^2_3}{4|h_2|}} \cos{[\sqrt{|h_2|}(t{-}t_{+})]}
\label{P4}
\end{equation}
has an infinite number of nulls, and changes the sign with the frequency $\sqrt{|h_2|}$. The periodic acceleration parameter
\begin{equation}
- q = 1- \frac{1}{h_+} \left\{\frac{\sin{[\sqrt{|h_2|}(t{-}t_{+})]}}{\cos^2{[\sqrt{|h_2|}(t{-}t_{+})]}} \right\}
\label{P5}
\end{equation}
signals that there are infinite number of epochs of deceleration and acceleration in the Universe evolution, and this parameter becomes infinite, when the Hubble function takes zero values.

\begin{figure}[H]
	\centering
	\includegraphics[width=\columnwidth]{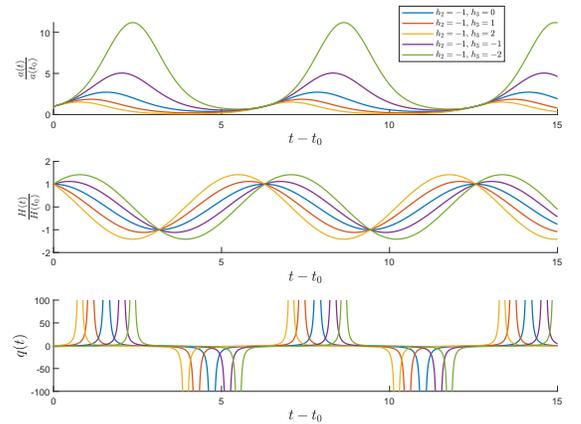}
	\caption{Illustration of the behavior of dimensionless scale factor $a(t)/a(t_0)$ (upper panel), Hubble functon $H(t)/H(t_0)$ (middle panel) and acceleration parameter $q(t)$ (bottom panel) as functions of cosmological time $t$ for the set of $h_3$ presented in the upper right corner of the figure; for all plots here we assume for simplicity $h_2 = -1$.}
	\label{fig3}
\end{figure}

The regime is illustrated in Fig.\ref{fig3}.

\subsubsection{$h_2=0$, the symmetric bounce}

In this particular case the state parameters of the system are linked by the relationship
\begin{equation}
E^{\prime}(1)= -  \frac{|\omega_0|}{(3-|\omega_0|)} W^{\prime}(1) \,,
\label{P58}
\end{equation}
and thus, the sign of the parameter
\begin{equation}
h_3 = - \kappa W^{\prime}(1) \left[\frac{2|\omega_0|-3}{3-|\omega_0|} \right] \approx - 1.16 \kappa W^{\prime}(1)
\label{P59}
\end{equation}
is predetermined by the sign of the initial value of the derivative $W^{\prime}(1)$.
We obtain now that the scale factor has the Gaussian form
\begin{equation}
a(t) = a(t_{*}) \exp\left[\frac14 h_3 (t-t_{*})^2 \right] \,,
\label{P6n}
\end{equation}
where the auxiliary parameters are
\begin{equation}
t_{*} = t_0 - 2 \frac{|H(1)|}{h_3} < t_0 \,,
\label{P7}
\end{equation}
\begin{equation}
a(t_{*}) = a(t_0) \exp\left(- \frac{H^2(1)}{h_3} \right) \,.
\label{P8}
\end{equation}
Clearly, the Universe expands, when $h_3>0$, i.e., $W^{\prime}(1)<0$ and $E^{\prime}(1)>0$.
For this solution the Hubble function is the linear function of the cosmological time
\begin{equation}
H(t) = \frac12 h_3 (t-t_{*}) \,,
\label{P9}
\end{equation}
and we deal with the solution, which at $h_3>0$ can be indicated as the symmetric bounce (see, e.g., \cite{Bounce1}).
The acceleration parameter
\begin{equation}
-q= 1+ \frac{2}{h_3(t-t_{*})^2}
\label{P10}
\end{equation}
is presented by the monotonic function at $h_3>0$, it tends to one asymptotically.

\subsection{Complex conjugated pair of roots}

\subsubsection{Solution to the key equation}

Let us consider the model, in which the characteristic equation is of the third order, one of the root is equal to zero, and two roots are complex conjugated with vanishing real parts.
For this model we obtain from (\ref{qq3})
\begin{equation}
\nu_{11} = {-} (3{+}2\omega_0) \,, \quad
 K^0_{11} =  \frac{ \omega_0 (3{+}2\omega_0)}{3+ \omega_0} \,.
 \label{CC1}
\end{equation}
The characteristic equation reduces now to
\begin{equation}
\sigma^2 + \beta^2 = 0 \,,
\label{CC2}
\end{equation}
where the quantity
\begin{equation}
\beta^2 = -  \frac{3 (3+\omega_0)^2  {+} \omega_0 (3+2\omega_0)^2}{3+\omega_0}
\label{CC3}
\end{equation}
is considered to be positive due to a special choice of the parameter $\omega_0$. (For instance, when $\omega_0= -\frac52$ and thus $\nu_{11}=2$, $K^0_{11}=10$, we obtain that $\beta^2 =  \frac{37}{2}> 0$). For the presented model we reconstruct the solution for the DE energy density in the form
\begin{equation}
W(x) = W(1) {+} \frac{W^{\prime}(1)}{\beta} \sin{(\beta \log{x})} +
\label{CC37}
\end{equation}
$$
+  \frac{1}{\beta^2}\left[W^{\prime \prime}(1) {+} W^{\prime}(1) \right] \left[1{-} \cos{(\beta \log{x})}\right] \,.
$$
According to (\ref{qq9}) the DM energy density can be recovered as
\begin{equation}
E(x)=  \frac{1}{3\omega_0 } \left[- \frac{x^2 W^{\prime \prime}(x)}{(2{+}\omega_0)} {+} x W^{\prime}(x) \right] + \frac{\omega_0 W(x)}{3{+}\omega_0} \,,
\label{qq96}
\end{equation}
and can be rewritten in the form similar to (\ref{CC37})
\begin{equation}\nonumber
E(x) = E(1)  {+} \frac{E^{\prime}(1)}{\beta} \sin{(\beta \log{x})}  +
\end{equation}
\begin{equation}\label{qq966}
+ \frac{1}{\beta^2} \left[E^{\prime \prime}(1) + E^{\prime}(1)\right]\left[1{-} \cos{(\beta \log{x})}\right] \,,
\end{equation}
where the initial values $E(1)$, $E^{\prime}(1)$ and $E^{\prime \prime}(1)$ are linked with  $W(1)$, $W^{\prime}(1)$ and $W^{\prime \prime}(1)$ as follows:
\begin{equation}
E(1) = \frac{1}{3\omega_0 } \left[- \frac{W^{\prime \prime}(1)}{(2{+}\omega_0)} {+} W^{\prime}(1) \right] + \frac{\omega_0 W(1)}{3{+}\omega_0} \,,
\label{qq94}
\end{equation}
\begin{equation}\nonumber
E^{\prime}(1) =  - \frac{(3+\omega_0)}{3\omega_0} \left[- \frac{W^{\prime \prime}(1)}{(2{+}\omega_0)} {+} W^{\prime}(1) \right] =
\end{equation}
\begin{equation}\label{2qq94}
= \omega_0 W(1) - (3+\omega_0) E(1) \,,
\end{equation}
\begin{equation}
E^{\prime \prime}(1) = {-} \frac{(3{+}\omega_0)}{3\omega_0 (2{+}\omega_0)}\left[W^{\prime \prime}(1) (4{+}\omega_0) {+} W^{\prime}(1)(1{+}\beta^2) \right]  \,.
\label{3qq94}
\end{equation}

\subsubsection{Geometric characteristics of the model}

The next step is to calculate the square of the Hubble function; again it can be represented in the form:
\begin{equation}
H^2(x) =  H^2(1) {+}  {\cal A}\left[1{-} \cos{(\beta \log{x})}\right] {+} {\cal B} \sin{(\beta \log{x})} \,,
\label{m1}
\end{equation}
where
\begin{equation}
H^2(1) = \frac{\Lambda}{3} + \frac{\kappa}{3} \left[W(1)+E(1) \right] \,,
\label{m0}
\end{equation}
\begin{equation}
{\cal A} = \frac{\kappa}{3\beta^2} \left[W^{\prime \prime}(1) + E^{\prime \prime}(1) + W^{\prime}(1) + E^{\prime}(1) \right] \,,
\label{m2}
\end{equation}
\begin{equation}
{\cal B} = \frac{\kappa}{3\beta} \left[W^{\prime}(1) + E^{\prime}(1) \right] \,.
\label{m3}
\end{equation}
Now the scale factor $a(t)$ can be found from the integral
\begin{equation}
\beta(t{-}t_0)  = \pm \int\limits^{\beta \log{\frac{a(t)}{a(t_0)}}}_{0} \frac{d z}{\sqrt{{\cal F} {-} {\cal A}\cos{z} {+} {\cal B} \sin{z}}} \,,
\label{m4}
\end{equation}
where we introduced the following auxiliary quantities
\begin{equation}
{\cal F} = H^2(1) +  {\cal A} \,,  \quad z = \beta \log{x} \,.
\label{m41}
\end{equation}
If we introduce the notations
\begin{equation}
\theta = \frac12 (z-z_{*}) \,, \quad \tan{z_{*}} = - \frac{{\cal B}}{\cal{A}} \,,
\label{m42}
\end{equation}
and assume that ${\cal F} {+} \sqrt{{\cal A}^2 {+} {\cal B}^2}>0$, we obtain
\begin{equation}\nonumber
\pm \frac12 \beta \sqrt{{\cal F} {+} \sqrt{{\cal A}^2 {+} {\cal B}^2}} \ (t{-}t_0) =
\end{equation}
\begin{equation}\label{m32}
=  \int\limits^{\frac12\left(\beta \log{\frac{a(t)}{a(t_0)}} - z_{*} \right)}_{-\frac12 z_{*}} \frac{d \theta}{\sqrt{1 {-} {\cal K}^2 \sin^2{\theta} }} \,,
\end{equation}
where
\begin{equation}
{\cal K}^2 = \frac{2 \sqrt{{\cal A}^2 {+} {\cal B}^2}}{{\cal F} {+} \sqrt{{\cal A}^2 + {\cal B}^2}} \,.
\label{m34}
\end{equation}
Clearly, we deal again with the incomplete elliptic integrals of the first kind and obtain
\begin{equation}
F\left(\frac12 \beta \log{\frac{a(t)}{a(t_0)}} - \frac12 z_{*}  | \  {\cal K}^2 \right) = {\cal J}(t) \,,
\label{m61}
\end{equation}
$$
{\cal J}(t)= F\left({-} \frac12 z_{*} | \ {\cal K}^2 \right) \pm \frac12 \beta \sqrt{{\cal F} {+} \sqrt{{\cal A}^2 {+} {\cal B}^2}} \ (t{-}t_0) \,.
$$
The scale factor can be expressed in terms of the inverse elliptic functions
\begin{equation}
a(t) = a(t_0) \exp\left\{\frac{1}{\beta}\left[ z_{*} + 2 am \left({\cal J}(t) | {\cal K}^2 \right)\right]\right\} \,.
\label{m69}
\end{equation}
The Hubble function is now quasi-periodic (see (\ref{m1})); its maximal and minimal values are predetermined by the inequalities
\begin{equation}
{\cal A} {-} \sqrt{{\cal A}^2 {+} {\cal B}^2} \leq  H^2(t) {-} H^2(1) \leq  {\cal A} {+} \sqrt{{\cal A}^2 {+} {\cal B}^2} \,.
\label{m86}
\end{equation}
The acceleration parameter
\begin{equation}
-q(x) = 1{+}\frac{\beta}{2H^2(x)} \left[{\cal A} \sin{(\beta \log{x})} {+}  {\cal B}\cos{(\beta \log{x})}\right]
\label{m88}
\end{equation}
is also quasi-periodic. Formally speaking, the function $q(t)$ can change the sign for special choice of the guiding parameters of the model and initial values of the DE and DM energy density scalars.

\section{Non-local cross-action of DE on DM: Examples of explicit representation of the solutions to the master equations}

\subsection{Exact solutions for the state functions}

We consider the model, for which three simplifications are assumed. First of all, we assume that $\omega_0=0$, i.e., the local link between DE and DM is absent. Second, we assume that DM does not act on DE, and thus, $K^0_{12} =0$ and $\nu_{12} =0$. Third, there exists the non-local cross action of DE on DM, but the self-interaction in the DM itself is absent, i.e., $K^0_{22} =0$, $\nu_{22} =0$.
For this case the scheme of derivation of the key equation is the following. We extract the pressure $P(x)$ from the conservation law for the DE
\begin{equation}
P(x) = - \frac13 \left[x W^{\prime} + 3W \right] \,,
\label{qKey2}
\end{equation}
and put it into the modified equation of state
\begin{equation}
x W^{\prime} + 3\Gamma W = -3 x^{-\nu_{11}}K_{11}^0 \int\limits_1^x dy y^{\nu_{11}-1}W(y)
 \,.
\label{qKey4}
\end{equation}
The differential version of this integral equation
\begin{equation}
x^2 W^{\prime \prime} + x W^{\prime} (1+\nu_{11} + 3\Gamma) + 3W (\Gamma \nu_{11} + K^0_{11}) = 0
\end{equation}
presents the Euler equation of the second order; the corresponding characteristic equation
\begin{equation}
\sigma^2 + \sigma(\nu_{11}+3\Gamma) + 3(\Gamma \nu_{11} + K^0_{11}) =0
\label{qqKey48}
\end{equation}
gives the roots
\begin{equation}
\sigma_{1,2} = - \frac12 (\nu_{11}+3\Gamma) \pm \sqrt{\frac14 (\nu_{11}-3\Gamma)^2 -3K^0_{11}} \,.
\end{equation}
Now one can obtain three different situations: there are two different real roots ($\sigma_1 \neq \sigma_2$), two coinciding real roots ($\sigma_1{=}\sigma_2$), and there is the pair of complex conjugated toots ($\sigma_{1,2} = \alpha \pm i \beta$). In all these cases the methods of the presentation of the solutions for $W(x)$ is well documented.

Then we extract the pressure $\Pi(x)$ from the  conservation law for the DM
\begin{equation}
\Pi(x) = - \frac13 \left(x E^{\prime} + 3 E \right) \,,
\end{equation}
and put it into the equation of state for DM; as the result we obtain the solution for $E(x)$ in quadratures
\begin{equation}\nonumber
E(x) = E(1) x^{-3\gamma} -
\end{equation}
\begin{equation}\label{qKey5}
{-}3K_{21}^0 x^{-3\gamma} \int\limits_1^x dz z^{3\gamma {-}\nu_{21} {-}1} \int\limits_1^z dy y^{\nu_{21}{-}1}W(y) \,.
\end{equation}
For all three variants of the structure of the function $W(x)$ the integral in (\ref{qKey5}) gives the solution for $E(x)$ in terms of elementary functions.

\subsection{Two real roots $\sigma_1 \neq \sigma_2$}

\subsubsection{The explicit solution}

This situation corresponds to the case, when the guiding parameters $\nu_{11}$, $\Gamma$ and $K^0_{11}$ satisfy the inequality
\begin{equation}
\frac14 (\nu_{11}-3\Gamma)^2 >3K^0_{11} \,.
\label{Ex1}
\end{equation}
Respectively, we obtain for the DE energy density
\begin{equation}
W(x) = C_1 x^{\sigma_1} + C_2 x^{\sigma_2} \,,
\label{Ex2}
\end{equation}
where the constants of integration are connected with the initial data as follows:
\begin{equation}
C_1 = \frac{W^{\prime}(1)-\sigma_2W(1)}{\sigma_1-\sigma_2} \,, \quad C_2 = \frac{- W^{\prime}(1)+\sigma_1W(1)}{\sigma_1-\sigma_2} \,.
\label{Ex3}
\end{equation}
If we put $W(x)$ from (\ref{Ex2}) into (\ref{qKey5}) the integration procedure gives the DM energy density in the form
\begin{equation}\nonumber
E(x) {=}  x^{{-}3\gamma} \left\{E(1) {+} \frac{3 K^0_{21}}{(\nu_{21}{-}3\gamma)} \left[\frac{C_1}{(\sigma_1{+}3\gamma)} {+} \frac{C_2}{(\sigma_2{+}3\gamma)}\right] \right\} {-}
\label{Ex4}
\end{equation}
\begin{equation}\nonumber
-3 K^0_{21} \left\{\frac{C_1 x^{\sigma_1}}{(\sigma_1{+}\nu_{21})(\sigma_1{+}3\gamma)} {+}  \frac{C_2 x^{\sigma_2}}{(\sigma_2{+}\nu_{21})(\sigma_2{+}3\gamma)} {+} \right.
\end{equation}
\begin{equation}
\left. + \frac{x^{-\nu_{21}}}{\nu_{21}-3\gamma}\left(\frac{C_1}{(\sigma_1+\nu_{21})} + \frac{C_2}{(\sigma_2+\nu_{21})} \right) \right\} \,.
\end{equation}
The square of the Hubble function can be now found as
\begin{equation}
H^2(x)  = \frac{\Lambda}{3} {+}\frac{\kappa}{3} \left[\Theta_1 x^{\sigma_1} {+} \Theta_2 x^{\sigma_2} {+}\Theta_3 x^{{-} \nu_{21}} {+}  \Theta_4 x^{{-}3\gamma}\right] \,,
\label{Ex6}
\end{equation}
where the following auxiliary parameters are introduced:
\begin{equation}
\Theta_1 = C_1 \left[1- \frac{3 K^0_{21}}{(\sigma_1+ \nu_{21})(\sigma_1+ 3\gamma)} \right] \,,
\label{Ex7}
\end{equation}
\begin{equation}
\Theta_2 = C_2 \left[1- \frac{3 K^0_{21}}{(\sigma_2+ \nu_{21})(\sigma_2+ 3\gamma)} \right] \,,
\label{Ex8}
\end{equation}
\begin{equation}
\Theta_3 = - \frac{3 K^0_{21}}{(\nu_{21}-3\gamma)}\left[\frac{C_1}{(\sigma_1+\nu_{21})} + \frac{C_2}{(\sigma_2+\nu_{21})} \right] \,,
\label{Ex9}
\end{equation}
\begin{equation}
\Theta_4 = E(1) + \frac{3 K^0_{21}}{(\nu_{21}{-}3\gamma)} \left[\frac{C_1}{(\sigma_1{+}3\gamma)} {+} \frac{C_2}{(\sigma_2{+}3\gamma)}\right] \,.
\label{Ex10}
\end{equation}
The acceleration parameter can be represented  in terms of $x$ as follows:
\begin{equation}\nonumber
-q(x) = 1 + \frac{\kappa}{6H^2}\left[\sigma_1 \Theta_1 x^{\sigma_1} {+} \sigma_2 \Theta_2 x^{\sigma_2} {-} \right.
\end{equation}
\begin{equation}\label{Ex11}
\left. -\nu_{21} \Theta_3 x^{{-} \nu_{21}} {-} 3\gamma \Theta_4 x^{{-}3\gamma} \right] \,.
\end{equation}

\subsubsection{The example of exact analysis}

One can see from (\ref{qqKey48}) that $\sigma_1$ and $\sigma_2$ have opposite sings, if $K^0_{11}<-\Gamma \nu_{11}$. Since one of the roots happens to be positive in this case, the DE energy density, the DM energy density and the square of the Hubble function  infinitely grow at $x \to \infty$; we omit this version of the theory.

When $K^0_{11}>-\Gamma \nu_{11}$, the parameters $\sigma_1$ and $\sigma_2$ are of the same sign; we assume that they are negative, obtaining the supplementary inequality $\nu_{11}>-3 \Gamma$. Now the functions $W(x)$ and $E(x)$ vanishes asymptotically, the Hubble function tends to the de Sitter value $H \to \sqrt{\frac{\Lambda}{3}}$, and $-q(x) \to 1$.

In case when $K^0_{11}=-\Gamma \nu_{11}$, one of the roots takes zero value, say, $\sigma_1=0$, and the second root $\sigma_2= -(\nu_{11}+ 3\Gamma)$ is again negative, if $\nu_{11}>-3 \Gamma$. For this submodel the state functions $W(x)$ and $E(x)$ tend asymptotically to their constant values
\begin{equation}
W(x) \to W_{\infty} = W(1) + \frac{W^{\prime}(1)}{(\nu_{11}+3\Gamma)} \,,
\label{Ex12}
\end{equation}
\begin{equation}
E(x) \to E_{\infty} = - \frac{K^0_{21} W_{\infty}}{\nu_{21}\gamma}\,.
\label{Ex121}
\end{equation}
We require that $E_{\infty}$ is positive and assume that $K^0_{21}<0$, $\nu_{21}>0$. As for the Hubble function, it tends to constant value
\begin{equation}
H(x) \to H_{\infty} = \sqrt{\frac{\Lambda}{3} + \frac{\kappa}{3}(W_{\infty}+E_{\infty})} \,,
\label{Ex13}
\end{equation}
providing the asymptotic regime to be of the de Sitter type.

Finally, in order to represent analytically the scale factor as the function of the cosmological time we  consider the following simple illustration. Let the
parameters of the model and the initial data be chosen specifically as follows
\begin{equation}
\sigma_1 = 0 \,, \quad 3K^0_{21} = (\sigma_2+\nu_{21})(\sigma_2+ 3\gamma) \,,
\label{Ex79}
\end{equation}
\begin{equation}
E(1) = \frac{W^{\prime}(1)}{3\gamma} \,, \quad W^{\prime}(1) = W(1)(\sigma_2+\nu_{21}) \,.
\label{Ex89}
\end{equation}
For this specific choice $\Theta_2 {=}\Theta_3 {=} \Theta_4 {=} 0$ and thus we obtain the constant Hubble function
\begin{equation}
H(x)  = H_{\infty} = \sqrt{\frac{\Lambda}{3} {+}\frac{\kappa W(1)}{9\gamma} \left[\nu_{21}{-}\nu_{11}{+} 3(\gamma-\Gamma) \right] } \,.
\label{Ex69}
\end{equation}
We deal with the example of solution describing the de Sitter type Universe.

\subsection{Two coinciding real roots $\sigma_1 = \sigma_2 \equiv \sigma $}

This situation corresponds to the case, when
\begin{equation}
\frac14 (\nu_{11}-3\Gamma)^2 =3K^0_{11} \,, \quad \sigma = - \frac12 (\nu_{11}+3\Gamma) \,,
\label{U1}
\end{equation}
Now  we obtain for the DE energy density
\begin{equation}
W(x) = x^{\sigma} \left[C_1  + C_2 \log{x}\right] \,,
\label{U2}
\end{equation}
where
\begin{equation}
C_1 = W(1) \,, \quad C_2 = W^{\prime}(1)-\sigma W(1)\,.
\label{U3}
\end{equation}
The DM energy density can be represented as
\begin{equation}\nonumber
E(x) =  x^{-3\gamma} \left[E(1) {+} \frac{3 K^0_{21}}{(\nu_{21}{-}3\gamma)(\sigma{+}3\gamma)} \left(C_1 {-} \frac{C_2}{\sigma{+}3\gamma}\right) \right]-
\end{equation}
\begin{equation}\nonumber
-3 K^0_{21} \left\{\frac{\left[C_1(\sigma{+}\nu_{21}){-}C_2 \right]}{(\sigma{+}\nu_{21})^2}\left[\frac{x^{\sigma}}{(\sigma{+}3\gamma)} {+}  \frac{x^{{-}\nu_{21}}}{(\nu_{21}{-}3\gamma)}\right] {+} \right.
\end{equation}
\begin{equation}\label{U4}
\left. + \frac{C_2 x^{\sigma} \left[(\sigma+3\gamma)\log{x}-1 \right]}{(\sigma+\nu_{21})(\sigma+3\gamma)^2} \right\} \,.
\end{equation}
The square of the Hubble function includes now the logarithm
\begin{equation}
H^2(x) = \frac{\Lambda}{3} + \frac{\kappa}{3} \left[x^{\sigma} \left(\tilde{\Theta}_1  {+}  \tilde{\Theta}_2 \log{x}\right) {+} \tilde{\Theta}_3 x^{{-} \nu_{21}} {+}   \tilde{\Theta}_4 x^{{-}3\gamma} \right] \,.
\label{U5}
\end{equation}
The new auxiliary functions are introduced as follows:
\begin{equation}\nonumber
\tilde{\Theta}_1 = C_1 \left[1- \frac{3 K^0_{21}}{(\sigma+ \nu_{21})(\sigma + 3\gamma)} \right] +
\end{equation}
\begin{equation}\label{U6}
+ C_2 \frac{3K_{21}(2\sigma+ 3\gamma+ \nu_{21})}{(\sigma+\nu_{21})^2(\sigma+3\gamma)^2} \,,
\end{equation}
\begin{equation}
\tilde{\Theta}_2 = C_2 \left[1- \frac{3 K^0_{21}}{(\sigma+ \nu_{21})(\sigma + 3\gamma)} \right] \,,
\label{U7}
\end{equation}
\begin{equation}
\tilde{\Theta}_3 = \frac{3 K^0_{21}}{(\nu_{21}{-}3\gamma)(\sigma{+}\nu_{21})^2}\left[C_2 {-}C_1 (\sigma{+}\nu_{21})\right] \,,
\label{U8}
\end{equation}
\begin{equation}
\tilde{\Theta}_4 = E(1) {+} \frac{3 K^0_{21}}{(\nu_{21}{-}3\gamma)(\sigma{+}3\gamma)} \left(C_1 {-} \frac{C_2}{\sigma{+}3\gamma}\right) \,.
\label{U9}
\end{equation}
The acceleration parameter is modified respectively
\begin{equation}\nonumber
-q(x) = 1 + \frac{\kappa}{6H^2}\left\{x^{\sigma} \left[\sigma \tilde{\Theta}_1  {+} \tilde{\Theta}_2 \left(\sigma \log{x}+1\right) \right] - \right.
\end{equation}
\begin{equation}\label{U10}
\left. -\nu_{21} \tilde{\Theta}_3 x^{{-} \nu_{21}} {-} 3\gamma \tilde{\Theta}_4 x^{{-}3\gamma} \right\} \,.
\end{equation}
Taking into account physical motives we assume that the parameter $\sigma$ is non-positive, i.e., $\nu_{11}+3\Gamma \geq0$.

\subsubsection{Illustration for the case $\sigma = 0$}

For illustration of an analytical result we assume that
\begin{equation}
W^{\prime}(1) = \nu_{21}W(1) \,, \quad E(1) = W(1)\frac{K^0_{21}}{3\gamma^2} \,,
\label{U11}
\end{equation}
providing the square of the Hubble function takes the simplified form
\begin{equation}
H^2(x) = \frac{\Lambda}{3} {+} \frac{\kappa}{3} \left(\tilde{\Theta}_1  {+}  \tilde{\Theta}_2 \log{x}\right) \,,
\label{U12}
\end{equation}
where
\begin{equation}
\tilde{\Theta}_1 = W(1)\left(1{+} \frac{K_{21}}{3\gamma^2} \right) \,, \quad  \tilde{\Theta}_2 = W(1) \left(\nu_{21}{-}\frac{K_{21}}{\gamma} \right) \,.
\label{U14}
\end{equation}
For this Hubble function the scale factor is of the symmetric bounce type \cite{Bounce1}
\begin{equation}
a(t)= a(t_*) e^{{\cal Q}(t-t_{*})^2} \,,
\label{U15}
\end{equation}
where
\begin{equation}
a(t_{*}) = a(t_0)) e^{- \frac{3H^2(1)}{\kappa \tilde{\Theta}_2}} \,, \quad {\cal Q} = \frac{\kappa \tilde{\Theta}_2}{12} \,.
\label{U16}
\end{equation}

\subsection{Complex conjugated roots $\sigma_{1,2} = \alpha \pm i\beta $}

This situation corresponds to the case, when
\begin{equation}
\frac14 (\nu_{11}-3\Gamma)^2 <3K^0_{11} \,,
\end{equation}
\begin{equation}\nonumber
\alpha = - \frac12 (\nu_{11}+3\Gamma) \,, \quad \beta = \sqrt{3K^0_{11} - \frac14 (\nu_{11}-3\Gamma)^2 } \,.
\label{U17}
\end{equation}
We obtain now
\begin{equation}\nonumber
W(x) = x^{\alpha} \left\{W(1) \cos{(\beta\log{x})} + \right.
\end{equation}
\begin{equation}\label{U19}
\left. + \frac{W^{\prime}(1)-\alpha W(1)}{\beta} \sin{(\beta\log{x})} \right\} \,.
\end{equation}
Clearly, the DE energy density changes the sign inevitably, so that the model seem to be non-physical.

\section{Discussion and conclusions}

\noindent
1) The main result of the presented work is the analysis of one specific rheologic-type model of interaction between the dark energy and dark matter. We introduce into the DE and DM equations of state four integral operators of the Volterra type; two of them describe the DE/DM cross-coupling, and two operators relate to the self-interactions in the DE and DM individually. The Volterra operators are chosen to correspond to the paradigm of fading memory, i.e., the kernels of these operators are of the difference type and multiplicative.

\noindent
2) The established model belongs to the class of exactly integrable models, i.e., the DE and DM state functions (energy densities and pressures), as well as, the Hubble function and the acceleration parameter are presented in the elementary functions. It has become possible since the key equation of the model is the linear Euler equation of the sixth order in ordinary derivatives.  The scale factor as the function of the cosmological time is found in quadratures and is studied analytically, qualitatively and numerically.

\noindent
3) The four-kernel model of the DE/DM interactions contains eight new guiding parameters; four of them, $\nu_{ij}$, describe the rates of memory fading, and other four $K^0_{ij}$ describe the effectiveness of the cross-coupling and self-interactions, respectively (see (\ref{K1})). On the one hand, such a multiparametricity extends the analytic possibilities for modeling of the Universe expansion. On the other hand, we use two instruments to constrain the set of these parameters. The first instrument is connected with the asymptotic analysis of the model; we assume that the Big Rip scenaria have to be avoided and require that the late-time Universe expansion is accelerated. As the result, we claim, for instance, that all the exact solutions corresponding to the roots of the characteristic equation, which have positive real parts, are non-physical. The second instrument relates to the requirement that the DE and DM energy densities scalars have to be positive during all the interval of the Universe evolution, thus imposing taboo for a few quasi-periodic regimes corresponding to the complex conjugated roots of the characteristic equation.

\noindent
4) In order to illustrate the general conclusions, we considered four examples of exact explicit solutions, which already appeared, e.g., in the framework of modified theories of gravity.
The first example describes the super-exponential (or super-inflationary) growth of the scale factor and exponential laws for the DE/DM state functions (see, (\ref{LR1}), (\ref{LR7})); this example belongs to the class of solutions of the Little Rip type. The second example relates to the solution known as the symmetric bounce (see (\ref{P6n}) and (\ref{U15})). The solutions of this type also belong to the class of the Little Rip from the point of view of asymptotic behavior; as for the global point of view, this solution is nonsingular, and the Hubble function is the linear function of time. The third example can be indicated as Pseudo Rip; the corresponding Hubble function tends asymptotically to constant $H_{\infty} \neq \sqrt{\frac{\Lambda}{3}}$ (see (\ref{Ex13})); the interesting feature of this solution is that the DE and DM energy densities tend asymptotically to nonvanishing constants (\ref{Ex12}), (\ref{Ex121}), and it is the explicit result of the non-local interactions. The fourth example relates to the solution with constant Hubble function  $H = {\rm const} \neq \sqrt{\frac{\Lambda}{3}}$ (see (\ref{Ex69})). One can indicate this solution as the  de Sitter type one; the Hubble constant includes now the rheological parameters.

\noindent
5)  We presented two exact explicit quasi-periodic solutions. For the first solution the  scale factor is of the form (\ref{P1}), the Hubble function is presented by the formula (\ref{P4}), the behavior of the DE energy density can be reconstructed using (\ref{qq34}). The frequency of oscillations is associated with the parameter $\sqrt{|h_2|}$ (\ref{qq42}), which is linked, formally
speaking, with the parameter of the local DE/DM interaction, $\omega_0$; however, the corresponding truncated model is obtained with the conditions (\ref{qq32}), which include the parameters of the non-local interaction. The second quasi-periodic solution is presented in terms of the incomplete elliptic integrals; the scale factor is presented by the formula (\ref{m69}) and the Hubble function can be correspondingly extracted from (\ref{m1}).

\acknowledgments{The work was supported by the Russian Science Foundation (Grant No 21-12-00130).}

\vspace{5mm}

\noindent
{\bf Appendix I:}

\noindent
{\bf Key equation for the case $\omega_0 \neq 0$}

\vspace{3mm}
Starting from the formulas
\begin{equation}\label{Enrel0}
 E(x) = \frac{1}{\omega_0}\left[x W^{\prime}(x) + (3+\omega_0)W + 3P \right] \,,
\end{equation}
$$
\Pi(x) = - \frac{1}{3\omega_0}\left\{x^2 W^{\prime \prime} + (7 + 2\omega_0)xW^{\prime} + (9 + 6\omega_0)W  + \right.
$$
\begin{equation}\label{Pirel0}
\left. +3xP^{\prime} + (9 + 3\omega_0)P \right\} \,,
\end{equation}
we obtain a pair of equations containing only the DE state functions, the pressure $P$ and the energy density $W$:
\begin{equation}
x^2P^{\prime \prime} + e x P^{\prime} + f P = bx^2 W^{\prime \prime} + cxW^{\prime} + dW \,,
\label{dif13}
\end{equation}
$$
{\cal E} xP^{\prime} {+} FP =
$$
\begin{equation}
= x^4W^{(IV)}{+}Ax^3 W^{(III)}{+}Bx^2 W^{\prime \prime}{+}CxW^{\prime}{+}DW  \,.
\label{dif3}
\end{equation}
The auxiliary parameters $e,f,b,c,d$  are written as follows:
$$
b = \Gamma-1 +\frac{K_{12}^0}{\omega_0} \,,
$$
$$
 c = (\Gamma{-}1)(\nu_{11}{+}\nu_{12}{+}1) {+} K_{11}^0 {+} \frac{K_{12}^0}{\omega_0}(4{+}\omega_0{+}\nu_{11})\,,
$$
$$
d = \nu_{11}\nu_{12}(\Gamma-1)+ K_{11}^0\nu_{12} +\frac{\nu_{11}K_{12}^0}{\omega_0}(3+\omega_0) \,,
$$
\begin{equation}
e = \nu_{11}{+}\nu_{12}{+}1{-}3\frac{K_{12}^0}{\omega_{0}} \,, \quad
f = \nu_{11}\nu_{12} {-} 3\frac{\nu_{11}K_{12}^0}{\omega_0} \,.
\end{equation}
Similarly we can represent the parameters ${\cal E},F,A,B,C,D$:
$$
A = \nu_{22}+\nu_{21}+6+2\omega_0+3\gamma+3\Gamma+\frac{3K_{12}^0}{\omega_0} \,,
$$
$$
B = 34+12\omega_0 +(\nu_{22}+\nu_{21})(9+2\omega_0)+\nu_{22}\nu_{21}+
$$
$$
+3(\gamma-1)(\nu_{22}+\nu_{21}+6+\omega_0)+3K_{22}^0+3(\Gamma-1)(\nu_{11}+\nu_{12}+3)  +
$$
$$
+ 3K_{11}^0 +\frac{3K_{12}^0}{\omega_0}(6+\omega_0+\nu_{11}) +
$$
$$
+3\left[\omega_0{+}(\nu_{22}{+}\nu_{21}){+}3\gamma {-} \left(\nu_{11}{+}\nu_{12}{-}3\frac{K_{12}^0}{\omega_0}\right)\right]\times
$$
$$
\times \left(\Gamma-1+\frac{K_{12}^0}{\omega_0}\right) \,,
$$
$$
C = (\nu_{22}+\nu_{21}+1)(16+8\omega_0)+\nu_{22}\nu_{21}(7+2\omega_0)+
$$
$$
+3\left[(\gamma-1)(\nu_{22}+\nu_{21}+1)+K_{22}^0\right](4+\omega_0)+
$$
$$
+3\left[\nu_{22}\nu_{21}(\gamma-1)+K_{22}^0\nu_{21}\right] +
$$
$$
+ 3\omega_0K_{21}^0 {+} 3(\Gamma{-}1)(\nu_{11}{+}\nu_{12} {+} \nu_{11}\nu_{12} {+}1){+}3K_{11}^0(1{+}\nu_{12}) +
$$
$$
+ \frac{3K_{12}^0}{\omega_0}(4{+}\omega_0+\nu_{11}) {+} 3\frac{K_{12}^0\nu_{11}}{\omega_0}(3{+}\omega_0) {+}
$$
$$
+ 3\left[\omega_0+(\nu_{22}{+}\nu_{21}){+}3\gamma - \left(\nu_{11}+\nu_{12}-3\frac{K_{12}^0}{\omega_0}\right)\right]\times
$$
$$
\times \left[(\Gamma{-}1)(\nu_{11}{+}\nu_{12}{+}1){+}K_{11}^0{+}\frac{K_{12}^0}{\omega_0}(4{+}\omega_0{+}\nu_{11})\right] \,,
$$
$$
 D=\nu_{22}\nu_{21}(9+6\omega_0)+
 3\omega_0\nu_{22}K_{21}^0 +
 $$
 $$
 + 3\left[\nu_{22}\nu_{21}(\gamma-1)+
 K_{22}^0\nu_{21}\right](3+\omega_0) +
 $$
$$
  + 3\left[\omega_0{+}(\nu_{22}{+}\nu_{21}){+}3\gamma {-} \left(\nu_{11}{+}\nu_{12}{-}3\frac{K_{12}^0}{\omega_0}\right)\right]\times
$$
$$
\times \left[\nu_{11}\nu_{12}(\Gamma-1)+K_{11}^0\nu_{12}+\frac{K_{12}^0\nu_{11}}{\omega_0}(3+\omega_0)\right] \,,
$$
 $$
 {\cal E} = 3(\nu_{11}+ 1)(\nu_{12}+1) - 9\frac{K_{12}^0(\nu_{11}+1)}{\omega_0} -
 $$
 $$
 - (\nu_{22}{+}\nu_{21}+1)(3{+}3\omega_0+9\gamma){-}3\nu_{22}\nu_{21}{-}9K_{22}^0 +
 $$
 $$
 3\left[\omega_0{+}(\nu_{22}{+}\nu_{21}){+}3\gamma {-}
 \left(\nu_{11}{+}\nu_{12}{-}\frac{K_{12}^0}{\omega_0}\right)\right]\times
 $$
 $$
 \times \left(\nu_{11}+\nu_{12}+1-3\frac{K_{12}^0}{\omega_0}\right)\,,
$$
$$
F = -\nu_{22}\nu_{21}(3\omega_0+9\gamma)-9K_{22}^0\nu_{21} +
$$
$$
+ 3\left[\omega_0{+}(\nu_{22}{+}\nu_{21}){+}3\gamma {-}\left(\nu_{11}{+}\nu_{12}{-}3\frac{K_{12}^0}{\omega_0}\right)\right]\times
$$
\begin{equation}
\times \left(\nu_{11}\nu_{12} - 3\frac{K_{12}^0\nu_{11}}{\omega_0}\right) \,.
\label{z12}
\end{equation}
The last step is to extract the DE pressure $P(x)$ from the pair of equations (\ref{dif13}), (\ref{dif3}); we obtain for the DE pressure the following relationship:
\begin{equation}\label{dif333}
\left(eF-F-f{\cal E}-\frac{F^2}{{\cal E}}\right) P(x) =
\end{equation}
$$
=x^5 W^{(V)} + \left(3 + A + e - \frac{F}{{\cal E}}\right)x^4 W^{(IV)}+
$$
$$
{+}\left(2A {+} B {+} eA {-} \frac{AF}{{\cal E}}\right)x^3 W^{\prime \prime \prime} +
$$
$$
+\left(B{+}C {+} eB {-}\frac{BF}{{\cal E}} {-} b{\cal E} \right)x^2 W^{\prime \prime} {+}
$$
$$
+ \left(D {+} eC {-} \frac{CF}{{\cal E}} {-} c{\cal E}\right)x W^{\prime} + \left(eD {-} D {-} \frac{DF}{{\cal E}} {-} d{\cal E}\right)W  \,,
$$
and put this $P(x)$ into  the equation (\ref{dif13}). As the result, we obtain the key equation
$$
x^6 W^{(VI)} + \omega_1 x^5 W^{(V)} + \omega_2 x^4 W^{(IV)} + \omega_3 x^3 W^{\prime \prime \prime} +
$$
\begin{equation}
+ \omega_4 x^2 W^{\prime \prime} + \omega_5 x W^{\prime} + \omega_6 W =0  \,,
\label{key1000}
\end{equation}
in which the following coefficients are introduced
$$
\omega_1 =8{+}A{+}e \,, \quad \omega_2= 12{+}6A{+}4e {+} B{+}eA {+} f \,,
$$
$$
\omega_3= 6A+4B+C+3eA + eB - b{\cal E} + fA \,,
$$
$$
\omega_4 = 2B{+}2C{+}D{+}2eB{-}2b{\cal E}{+}eC{-}c{\cal E}{-}bF{+}fB \,,
$$
\begin{equation}
\omega_5 {=} eC{-}c{\cal E}{+}eD{-}d{\cal E}{-}cF{+}fC\,, \quad
\omega_6 {=} fD{-}dF \,.
\label{key101}
\end{equation}



\vspace{5mm}
\noindent
{\bf Appendix II:}

\noindent{\bf Key equation for the case $\omega_0 {=} 0$, $K^0_{12}\neq 0$, $K^0_{21}\neq 0$}

\vspace{3mm}
 When  $\omega_0 =0$, we extract the DE pressure $P$ from (\ref{Key1}) and the DM pressure from (\ref{Key2})
\begin{equation}
P = - \frac13 x W^{\prime} - W \,,
\quad
\Pi = - \frac13 x E^{\prime} - E \,.
\label{Key29m}
\end{equation}
Then we put these $P$ and $\Pi$ into the equations (\ref{dif1}) and (\ref{dif2}) obtaining two equations, which link now the DE and DM energy densities $W$ and $E$:
\begin{equation}
x^3 W^{\prime \prime \prime} {+} \alpha_1 x^2 W^{\prime \prime}  {+}
 \alpha_2 x W^{\prime}  {+} \alpha_3 W = {-} 3 K^0_{12}\left(xE^{\prime} {+} \nu_{11} E \right) \,,
\label{Key51}
\end{equation}
\begin{equation}
x^3 E^{\prime \prime \prime} {+} \alpha_4 x^2 E^{\prime \prime}  {+}
\alpha_5 x E^{\prime}  {+}  \alpha_6 E  = {-} 3 K^0_{21}\left(xW^{\prime} {+} \nu_{22} W \right) \,,
\label{Key52}
\end{equation}
where the new auxiliary parameters are the following:
$$
\alpha_1 = \left(3+ 3\Gamma + \nu_{11} + \nu_{12} \right) \,,
$$
$$
\alpha_2 = 3K^0_{11} + \left(1+3\Gamma \right)\left(1+ \nu_{11}+ \nu_{12} \right) \,,
$$
$$
\alpha_3 = 3 \nu_{12} \left(\nu_{11}\Gamma + K^0_{11} \right)
\quad
\alpha_4 = \left(3+ 3\gamma + \nu_{22} + \nu_{21} \right) \,,
$$
$$
\alpha_5 = 3K^0_{22} + \left(1+3\gamma \right)\left(1+ \nu_{22}+ \nu_{21} \right) \,,
$$
\begin{equation}
\alpha_6 = 3 \nu_{21} \left(\nu_{22}\gamma + K^0_{22} \right)\,.
\end{equation}
When $K^0_{12}\neq 0$, we find subsequently $E^{\prime \prime \prime}(x)$, $E^{\prime \prime}$, $E^{\prime}$ and $E$ from this pair of equations. For the DM energy density $E$ we obtain
\begin{equation} \label{EW0}
E(x) * 3K^0_{12} \left[\alpha_6 {+} (\alpha_4{-}2{-}\nu_{11})(1{+}\nu_{11})\nu_{11} {-}\alpha_5\nu_{11} \right] =
\end{equation}
$$
= x^5 W^{(V)} {+} x^4 W^{(IV)}\left(4 {+} \alpha_1 {+} \alpha_4 {-} \nu_{11}\right) +
$$
$$
+ x^3 W^{\prime \prime \prime} \left[6 {+} 4\alpha_1 {+} \alpha_2 {+} \alpha_5 {+} (\alpha_4 {-} 2 {-} \nu_{11}) (2{+}\alpha_1{-}\nu_{11}) \right] +
$$
$$
{+}x^2 W^{\prime \prime} \left[2\alpha_1 \nu_{11} {+} \alpha_3 {+} \alpha_1\alpha_5 {+} (\alpha_4 {-} \nu_{11})(\alpha_1 {+} \alpha_2 {-} \alpha_1\nu_{11})\right]
$$
$$
+x W^{\prime} \left[\alpha_2\alpha_5 {+} (\alpha_4 {-} 2 {-} \nu_{11})(\alpha_3 {-} \alpha_2\nu_{11})-9K^0_{12}K^0_{21}\right] +
$$
$$
+ W \left[\alpha_3\alpha_5 {-} \alpha_3(\alpha_4 {-} 2 {-} \nu_{11})(1+\nu_{11}) - 9K^0_{12}K^0_{21}\nu_{22}\right] \,.
$$
Then we put $E^{\prime \prime \prime}(x)$, $E^{\prime \prime}$, $E^{\prime}$ and $E$ into (\ref{Key52}) and find the key equation  of the sixth order in derivatives:
$$
x^6 W^{(VI)} {+} \Omega_1 x^5 W^{(V)} {+} \Omega_2 x^4 W^{(IV)} {+} \Omega_3 x^3 W^{\prime \prime \prime} {+}
$$
\begin{equation}
\Omega_4 x^2 W^{\prime \prime} + \Omega_5 x W^{\prime} + \Omega_6  W =0  \,,
\label{key10001}
\end{equation}
where the auxiliary parameters $\Omega_j$ are of the form
$$
\Omega_1 = 9 + \alpha_1 + \alpha_4 \,,
$$
$$
\Omega_2 = 30 + 8 \alpha_1 + \alpha_2 + \alpha_5 + (\alpha_4 - 2)(6 + \alpha_1) \,,
$$
$$
\Omega_3 = 18 {+} 14\alpha_1 {+} 5\alpha_2 {+} \alpha_3 {+} 3\alpha_5 {+} \alpha_6 {+} \alpha_1\alpha_5 +
$$
$$
{+} (\alpha_4 {-} 2)(6 {+} 4\alpha_1 {+} \alpha_2) \,,
$$
$$
\Omega_4 = 4\alpha_1 {+} 4\alpha_2 {+} 2\alpha_3 {+} 2\alpha_1\alpha_5 {+} \alpha_2\alpha_5 {+} \alpha_1\alpha_6 {+}
$$
$$
{+} (\alpha_4 {-} 2)(2\alpha_1 {+} 2\alpha_2 {+} \alpha_3) - 9K^0_{12}K^0_{21} \,,
$$
$$
\Omega_5 = \alpha_2\alpha_5 {+} \alpha_3\alpha_5 {+} \alpha_2\alpha_6 {-} 9K^0_{12}K^0_{21}(1+\nu_{11}+\nu_{22}) \,,
$$
\begin{equation}
\Omega_6 = \alpha_3\alpha_6 - 9K^0_{12}K^0_{21}\nu_{11}\nu_{22} \,.
\label{key1999}
\end{equation}

\end{document}